\pdfoutput=1

\documentclass[reprint,
superscriptaddress,
nobibnotes,
 amsmath,amssymb,
 aps,
prb,
citeautoscript
]{revtex4-1}
\usepackage{soul} 
\usepackage{ifthen}

\usepackage{bm}

\usepackage{tensor}
\usepackage{subfig}
\usepackage{braket}
\usepackage[export]{adjustbox}

\usepackage{nicefrac}
\usepackage{array}
\usepackage{multirow}

\usepackage[usenames, dvipsnames]{color}
\usepackage{graphicx}
\usepackage{booktabs}

\DeclareCaptionLabelFormat{andtable}{#1~#2  \&  \tablename~\thetable}
\newcommand{\myfont}[1]{\mbox{\textsc{#1}}}
\newcommand{\Eqref}[1]{Eq.~(\ref{#1})}

\newcommand{\Secref}[1]{Sec.~\ref{#1}}

\DeclareMathAlphabet\mathbfcal{OMS}{cmsy}{b}{n}

\newcommand{\partd}[2]{\ifmmode \frac{\partial #1}{\partial #2} \else $\frac{\partial #1}{\partial #2}$~\fi}
\newcommand{\partdd}[3]{\ifmmode \frac{\partial^2 #1}{\partial #2 \partial #3} \else $\frac{\partial^2 #1}{\partial #2 \partial #3}$~\fi}

\newcommand{\fulld}{\mathrm{d}}
\newcommand{\totald}[2]{\ifmmode \frac{\fulld #1}{ \fulld #2} \else $\frac{\fulld #1}{ \fulld #2}$~\fi}
\newcommand{\totaldd}[2]{\ifmmode \frac{\fulld^2 #1}{ \fulld #2^2} \else $\frac{\fulld^2 #1}{ \fulld #2^2}$~\fi}

\newcommand{\funcd}[2]{\ifmmode \frac{\delta #1}{ \delta #2} \else $\frac{\delta #1}{ \delta #2}$~\fi}
\newcommand{\funcdd}[2]{\ifmmode \frac{\delta^2 #1}{ \delta #2^2} \else $\frac{\delta^2 #1}{ \delta #2^2}$~\fi}

\newcommand{\bfr}{\ifmmode  \mathbf{r}  \else $ \mathbf{r} $~\fi}
\newcommand{\bfrp}{\ifmmode \mathbf{r}' \else $ \mathbf{r}' $~\fi}
\newcommand{\bfR}{\ifmmode  \mathbf{R}  \else $ \mathbf{R} $~\fi}
\newcommand{\bfa}[1]{\ifmmode \mathbf{a}_{#1} \else $\mathbf{a}_{#1}$~\fi}
\newcommand{\bfb}{\ifmmode \mathbf{b} \else $\mathbf{b} $~\fi}
\newcommand{\bfk}{\ifmmode  \mathbf{k}  \else $ \mathbf{k} $~\fi}
\newcommand{\bfkp}{\ifmmode  \mathbf{k}' \else $ \mathbf{k}' $~\fi}
\newcommand{\bfx}{\ifmmode \mathbf{x} \else $ \mathbf{x} $~\fi}
\newcommand{\bfxp}{\ifmmode \mathbf{x}' \else $ \mathbf{x}' $~\fi}
\newcommand{\bfp}{\ifmmode \mathbf{p} \else $ \mathbf{p} $~\fi}
\newcommand{\bfpp}{\ifmmode \mathbf{p}' \else $ \mathbf{p}' $~\fi}
\newcommand{\bfq}{\ifmmode \mathbf{q} \else $ \mathbf{q} $~\fi}
\newcommand{\bfv}{\ifmmode \mathbf{v} \else $ \mathbf{v} $~\fi}
\newcommand{\bfG}{\ifmmode \mathbf{G} \else $ \mathbf{G} $~\fi}
\newcommand{\bGamma}{\ifmmode \boldsymbol{\Gamma} \else $ \boldsymbol{\Gamma} $~\fi}
\newcommand{\btaup}{\ifmmode \boldsymbol{\tau}' \else $ \boldsymbol{\tau}' $~\fi}
\newcommand{\bfO}{\ifmmode \mathbf{0} \else $ \mathbf{0} $~\fi}

\newcommand{\bfe}{\ifmmode \mathbf{e} \else $\mathbf{e}$~\fi}

\newcommand{\difft}{\ifmmode \fulld t \else $ \fulld t $~\fi}
\newcommand{\diffr}{\ifmmode \fulld\bfr \else $ \fulld\bfr $~\fi}
\newcommand{\diffrp}{\ifmmode  \fulld\bfrp \else $ \fulld\bfrp $~\fi} 
\newcommand{\diffk}{\ifmmode \fulld\bfk \else $ \fulld\bfk $~\fi}
\newcommand{\diffkp}{\ifmmode \fulld\bfkp \else $ \fulld\bfkp $~\fi}
\newcommand{\diffx}{\ifmmode \fulld\bfx \else $ \fulld\bfx $~\fi}
\newcommand{\diffp}{\ifmmode \fulld\bfp \else $ \fulld\bfp $~\fi}
\newcommand{\diffpp}{\ifmmode \fulld\bfpp \else $ \fulld\bfpp $~\fi}
\newcommand{\diffR}{\ifmmode \fulld\bfR \else $ \fulld\bfR $~\fi}
\newcommand{\diffRN}{\ifmmode \fulld\bfR_1\cdots\fulld\bfR_N \else $ \fulld\bfR_1\cdots\fulld\bfR_N $~\fi}
\newcommand{\diffP}{\ifmmode \fulld\bfP \else $ \fulld\bfP $~\fi}
\newcommand{\diffPN}{\ifmmode \fulld\bfP_1\cdots\fulld\bfP_N \else $ \fulld\bfP_1\cdots\fulld\bfP_N $~\fi}

\newcommand{\intBZ}[1]{\int_{\mathrm{BZ}}\!\!#1\,}

\newcommand{\norm}[1]{\ifmmode \| #1 \| \else $\| #1 \|$~\fi}
\newcommand{\normsqd}[1]{\ifmmode \| #1 \|^2 \else $\| #1 \|^2$~\fi}
\newcommand{\modulus}[1]{\ifmmode \left\vert #1 \right\vert \else $\left\vert #1 \right\vert$~\fi}
\newcommand{\modulussqd}[1]{\ifmmode \left\vert #1 \right\vert^2 \else $\left\vert #1 \right\vert^2$~\fi}

\newcommand{\psir}[1]{\ifmmode     \psi_{#1}(\bfr)       \else $ \psi_{#1}(\bfr)      $~\fi}
\newcommand{\psiastr}[1]{\ifmmode \psi_{#1}^\ast(\bfr) \else $ \psi_{#1}^\ast(\bfr) $~\fi}
\newcommand{\psirp}[1]{\ifmmode    \psi_{#1}(\bfrp)      \else $ \psi_{#1}(\bfrp)     $~\fi}
\newcommand{\psiastpr}[1]{\ifmmode \psi_{#1}^\ast(\bfrp) \else $ \psi_{#1}^\ast(\bfrp) $~\fi}
\newcommand{\phir}[1]{\ifmmode     \phi_{#1}(\bfr)       \else $ \phi_{#1}(\bfr)      $~\fi}
\newcommand{\phirp}[1]{\ifmmode    \phi_{#1}(\bfrp)      \else $ \phi_{#1}(\bfrp)     $~\fi}
\newcommand{\phiastrp}[1]{\ifmmode \phi_{#1}^\ast(\bfrp) \else $ \phi_{#1}^\ast(\bfrp) $~\fi}
\newcommand{\phiastr}[1]{\ifmmode  \phi_{#1}^\ast(\bfr)  \else $ \phi_{#1}^\ast(\bfr) $~\fi}
\newcommand{\phia}{\ifmmode \phi_\alpha \else $ \upphi_\alpha $~\fi}
\newcommand{\phib}{\ifmmode \phi_\beta \else $ \upphi_\beta $~\fi}
\newcommand{\psink}[1]{\ifmmode  \psi_{#1\bfk}(\bfr)  \else $\psi_{#1\bfk}(\bfr)$~\fi}
\newcommand{\kunk}{\ifmmode \ket{u_{n\bfk}} \else $\ket{u_{n\bfk}}$~\fi}
\newcommand{\kumk}{\ifmmode \ket{u_{m\bfk}} \else $\ket{u_{m\bfk}}$~\fi}
\newcommand{\unk}{\ifmmode u_{n\bfk} \else $u_{n\bfk} $~\fi}
\newcommand{\umk}{\ifmmode u_{m\bfk} \else $u_{m\bfk} $~\fi}
\newcommand{\unkp}{\ifmmode u_{n\bfkp}(\bfr) \else $u_{n\bfkp}(\bfr)$~\fi}
\newcommand{\unkt}{\ifmmode \widetilde{u}_{n\bfk}(\bfr) \else $\widetilde{u}_{n\bfk}(\bfr)$~\fi}
\newcommand{\unkR}{\ifmmode u_{n\bfk}(\bfr+\bfR) \else $u_{n\bfk}(\bfr+\bfR)$~\fi}
\newcommand{\wannr}[2]{\ifmmode w_{#1#2}(\bfr) \else $ w_{#1#2}(\bfr) $~\fi}
\newcommand{\wann}[2]{\ifmmode w_{#1#2} \else $ w_{#1#2} $~\fi}
\newcommand{\kwann}[2]{\ifmmode \Ket{w_{#1 #2}} \else $ \Ket{w_{#1 #2}} $~\fi}
\newcommand{\kpsi}[2]{\ifmmode \Ket{\psi_{#1 #2}} \else $ \Ket{\psi_{#1 #2}} $~\fi}
\newcommand{\bpsi}[2]{\ifmmode \Bra{\psi_{#1 #2}} \else $ \Bra{\psi_{#1 #2}} $~\fi}

\newcommand{\omi}{\Omega_{\mathrm{I}}}
\newcommand{\omt}{\widetilde{\Omega}}
\newcommand{\omd}{\Omega_{\mathrm{D}}}
\newcommand{\omod}{\Omega_{\mathrm{OD}}}
\newcommand{\omiod}{\Omega_{\mathrm{IOD}}}
\newcommand{\ketbra}[2]{\ifmmode \ket{#1}\bra{#2} \else $ \ket{#1}\bra{#2} $~\fi}
\newcommand{\Ketbra}[2]{\ifmmode \Ket{#1}\Bra{#2} \else $ \Ket{#1}\Bra{#2} $~\fi}

\newcommand{\commutator}[2]{\ifmmode \left[#1,#2\right] \else $ \left[#1,#2\right] $~\fi}

\newcommand{\vecspace}[1]{\ifmmode \mathcal{#1} \else $ \mathcal{#1} $~\fi}
\newcommand{\cspace}{\ifmmode \mathbb{C} \else $ \mathbb{C} $~\fi}
\newcommand{\realspace}{\ifmmode \mathbb{R} \else $ \mathbb{R} $~\fi}

\newcommand{\elm}[3]{\ifmmode \braket{#1 \vert #2 \vert #3} \else $ \braket{ #1 \vert #2 \vert #3 } $~\fi }
\newcommand{\Elm}[3]{\ifmmode \Braket{#1 \vert #2 \vert #3} \else $ \Braket{ #1 \vert #2 \vert #3 } $~\fi }
\newcommand{\inprod}[2]{\ifmmode \Braket{ #1 | #2 } \else $ \Braket{ #1 \vert #2 } $~\fi} 
\newcommand{\overlap}[2]{\ifmmode \inprod{\phir{#1}}{\phir{#2}} \else $\inprod{\phir{#1}}{\phir{#2}} $~\fi} 

\newcommand{\sub}[1]{\ifmmode _{\mathrm{#1}} \else $_{\mathrm{#1}} $~\fi}
\newcommand{\super}[1]{\ifmmode ^{\mathrm{#1}} \else $ ^{\mathrm{#1}} $~\fi}
\newcommand{\mrm}[1]{\ifmmode   \mathrm{#1} \else $ \mathrm{#1} $~\fi}
\newcommand{\tinysub}[1]{\ifmmode _{\mbox{\scriptsize{#1}}} \else $ _{\mbox{\scriptsize{#1}}} $~\fi}
\newcommand{\tinysup}[1]{\ifmmode ^{\mbox{\scriptsize{#1}}} \else $ ^{\mbox{\scriptsize{#1}}} $~\fi}

\newcommand{\orthomat}[1]{\ifmmode \mathbf{#1}^{\perp} \else $ \mathbf{#1}^{\perp} $~\fi}
\newcommand{\trans}{\ifmmode \mathsf{T} \else $ \mathsf{T} $~\fi}
\newcommand{\transpose}[1]{\ifmmode {#1}^\trans \else $ {#1}^\trans $~\fi}
\newcommand{\mmn}{\ifmmode M_{mn}^{(\mathbf{k,b})} \else  $ M_{mn}^{(\mathbf{k,b})} $~\fi}
\newcommand{\mnn}{\ifmmode M_{nn}^{(\mathbf{k,b})} \else  $ M_{nn}^{(\mathbf{k,b})} $~\fi}
\newcommand{\amn}{\ifmmode A_{mn\mathbf{k}} \else $ A_{mn\mathbf{k}} $~\fi}
\newcommand{\smn}{\ifmmode S_{mn\mathbf{k}} \else $ S_{mn\mathbf{k}} $~\fi}
\newcommand{\umn}{\ifmmode U_{mn\mathbf{k}} \else $ U_{mn\mathbf{k}} $~\fi}
\newcommand{\vmn}{\ifmmode V_{mn\mathbf{k}} \else $ V_{mn\mathbf{k}} $~\fi}
\newcommand{\wmn}{\ifmmode W_{mn}^{(\mathbf{k})} \else $ W_{mn}^{(\mathbf{k})} $~\fi}
\newcommand{\dwmn}{\ifmmode \fulld W_{mn}^{(\mathbf{k})} \else $ \fulld W_{mn}^{(\mathbf{k})} $~\fi}
\newcommand{\zmn}{\ifmmode Z_{mn} \else $ Z_{mn} $~\fi}
\newcommand{\matUk}{\ifmmode U_{\bfk} \else $ U_{\bfk} $~\fi}
\newcommand{\matAk}{\ifmmode A_{\bfk} \else $ A_{\bfk} $~\fi}
\newcommand{\matSk}{\ifmmode S_{\bfk} \else $ S_{\bfk} $~\fi}
\newcommand{\matMkb}{\ifmmode M^{(\bfk,\bfb)} \else $ M^{(\bfk,\bfb)} $~\fi}
\newcommand{\matVk}{\ifmmode V_{\bfk} \else $ V_{\bfk} $~\fi}
\newcommand{\rmnkb}{\ifmmode R_{mn}^{(\mathbf{k,b})} \else  $ R_{mn}^{(\mathbf{k,b})} $~\fi}
\newcommand{\tmnkb}{\ifmmode T_{mn}^{(\mathbf{k,b})} \else  $ T_{mn}^{(\mathbf{k,b})} $~\fi}
\newcommand{\qnkb}{\ifmmode q_{n}^{(\mathbf{k,b})} \else  $ q_{n}^{(\mathbf{k,b})} $~\fi}
\newcommand{\matcalDk}{\ifmmode \mathcal{D}_{\bfk} \else $ \mathcal{D}_{\bfk} $~\fi}
\newcommand{\matcalGk}{\ifmmode \mathcal{G}_{\bfk} \else $ \mathcal{G}_{\bfk} $~\fi}

\newcommand{\trace}{\ifmmode \mathrm{tr} \else $ \mathrm{tr} $~\fi}
\newcommand{\Trace}{\ifmmode \mathrm{Tr} \else $ \mathrm{Tr} $~\fi}

\newcommand{\ON}[1]{\ifmmode \mathcal{O}\left(N^{#1}\right) \else $\mathcal{O}\left(N^{#1}\right)$~\fi}

\newcommand{\overbar}[1]{\mkern1.5mu\overline{\mkern-1.5mu#1\mkern-1.5mu}\mkern 1.5mu}

\newcommand{\eg}{e.g.}
\newcommand{\ie}{i.e.}

\newcommand{\Wannier}{\myfont{Wannier90}}
\newcommand{\aiidawan}{AiiDA-\Wannier{}}
\newcommand{\QE}{\myfont{Quantum ESPRESSO}}
\newcommand{\pwscf}{\myfont{pwscf}}

\newcommand{\MLWF}{\mbox{MLWF}}
\newcommand{\MLWFs}{\mbox{MLWFs}}
\newcommand{\SLWFs}{\mbox{SLWFs}}

\newcommand{\inputvar}[1]{{\tt #1}}
\newcommand{\module}[1]{{\tt #1.F90}}
\newcommand{\exec}[1]{{\tt #1.x}}
\newcommand{\package}[1]{\myfont{#1}}
\newcommand{\pwtowan}{{\exec{pw2wannier90}}}

\usepackage{hyperref}
\begin{document}

\title{Wannier90 as a community code: new features and applications}

\newcommand\WDG{Members of the Wannier Developers Group, who are responsible for the long-term maintenance and sustainability of \Wannier.}

\author{Giovanni Pizzi}
\altaffiliation{\WDG}
\affiliation{Theory and Simulation of Materials (THEOS) and National Centre for Computational Design and Discovery of Novel Materials (MARVEL), \'Ecole Polytechnique F\'ed\'erale de Lausanne (EPFL), CH-1015 Lausanne, Switzerland}
\author{Valerio Vitale}
\altaffiliation{\WDG}
\affiliation{Cavendish Laboratory, Department of Physics, University of Cambridge, Cambridge CB3 0HE, UK}\affiliation{Departments of Materials and Physics, and the Thomas Young Centre for Theory and Simulation of Materials, Imperial College London, London SW7 2AZ, UK}
\author{Ryotaro Arita}\affiliation{RIKEN Center for Emergent Matter Science, 2-1 Hirosawa, Wako, Saitama 351-0198, Japan}\affiliation{Department of Applied Physics, The University of Tokyo, 7-3-1 Hongo, Bunkyo-ku, Tokyo 113-8656, Japan}
\author{Stefan Bl\"ugel}\affiliation{Peter Gr\"unberg Institut and Institute for Advanced Simulation, Forschungszentrum J\"ulich and JARA, 52425 J\"ulich, Germany}
\author{Frank Freimuth}\affiliation{Peter Gr\"unberg Institut and Institute for Advanced Simulation, Forschungszentrum J\"ulich and JARA, 52425 J\"ulich, Germany}
\author{Guillaume G\'eranton}\affiliation{Peter Gr\"unberg Institut and Institute for Advanced Simulation, Forschungszentrum J\"ulich and JARA, 52425 J\"ulich, Germany}
\author{Marco Gibertini}\affiliation{Theory and Simulation of Materials (THEOS) and National Centre for Computational Design and Discovery of Novel Materials (MARVEL), \'Ecole Polytechnique F\'ed\'erale de Lausanne (EPFL), CH-1015 Lausanne, Switzerland}\affiliation{Department of Quantum Matter Physics, University of Geneva, Geneva, Switzerland}
\author{Dominik Gresch}\affiliation{ETH Zurich, Zurich, Switzerland}
\author{Charles Johnson}\affiliation{Departments of Materials and Physics, Imperial College London, London SW7 2AZ, UK}
\author{Takashi Koretsune}\affiliation{Department of Physics, Tohoku University, Sendai, Japan}\affiliation{JST PRESTO, Kawaguchi, Saitama, Japan}
\author{Julen Iba\~nez-Azpiroz}\affiliation{Centro de F\'isica de Materiales, Universidad del Pa\'is Vasco, E-20018 San Sebasti\'an, Spain}
\author{Hyungjun Lee}\affiliation{Institute of Physics, \'Ecole Polytechnique F\'ed\'erale de Lausanne (EPFL), CH-1015 Lausanne, Switzerland}
\affiliation{Korea Institute for Advanced Study, Hoegiro 85, Seoul 02455, Korea}
\author{Jae-Mo Lihm}
\affiliation{Department of Physics and Center for Theoretical Physics, Seoul National University, Seoul 08826, Korea}
\author{Daniel Marchand}\affiliation{Laboratory for Multiscale Mechanics Modeling (LAMMM), \'Ecole Polytechnique F\'ed\'erale de Lausanne (EPFL), Lausanne, Switzerland}
\author{Antimo Marrazzo}\affiliation{Theory and Simulation of Materials (THEOS) and National Centre for Computational Design and Discovery of Novel Materials (MARVEL), \'Ecole Polytechnique F\'ed\'erale de Lausanne (EPFL), CH-1015 Lausanne, Switzerland}
\author{Yuriy Mokrousov}\affiliation{Peter Gr\"unberg Institut and Institute for Advanced Simulation, Forschungszentrum J\"ulich and JARA, 52425 J\"ulich, Germany}
\affiliation{Institute of Physics, Jonannes-Gutenberg University of Mainz, 55099 Mainz, Germany}
\author{Jamal I. Mustafa}\affiliation{Department of Physics, University of California at Berkeley, Berkeley, California 94720, USA}
\author{Yoshiro Nohara}\affiliation{ASMS Co., Ltd., 1-7-11 Higashi-Gotanda, Shinagawa-ku, Tokyo 141-0022, Japan}
\author{Yusuke Nomura}\affiliation{RIKEN Center for Emergent Matter Science, 2-1 Hirosawa, Wako, Saitama 351-0198, Japan}
\author{Lorenzo Paulatto}\affiliation{Institut de Min\'eralogie, de Physique des Mat\'eriaux et de Cosmochimie (IMPMC), Sorbonne Universit\'e, CNRS UMR
7590, Case 115, 4 place Jussieu, 75252 Paris Cedex 05, France} 
\author{Samuel Ponc\'e}\affiliation{Department of Materials, University of Oxford, Parks Road, Oxford OX1 3PH, UK}
\author{Thomas Ponweiser}\affiliation{Research Institute for Symbolic Computation (RISC), Johannes Kepler University, Altenberger Stra{\ss}e 69, 4040 Linz, Austria}
\author{Junfeng Qiao}\affiliation{Fert Beijing Institute, School of Microelectronics, BDBC, Beihang University, Beijing, China}
\author{Florian Th\"ole}\affiliation{Materials Theory, ETH Z\"urich, Wolfgang-Pauli-Strasse 27, CH-8093 Z\"urich, Switzerland}
\author{Stepan S. Tsirkin}\affiliation{Centro de F\'isica de Materiales, Universidad del Pa\'is Vasco, E-20018 San Sebasti\'an, Spain}
\affiliation{Department of Physics, University of Zurich,
Wintherthurerstrasse 190, CH-8057 Zurich, Switzerland}
\author{Ma\l{}gorzata Wierzbowska}\affiliation{Institute of Physics, Polish Academy of Science, Al. Lotnik\'ow 32/46, 02-668 Warsaw, Poland}
\author{Nicola Marzari}
\altaffiliation{\WDG}
\affiliation{Theory and Simulation of Materials (THEOS) and National Centre for Computational Design and Discovery of Novel Materials (MARVEL), \'Ecole Polytechnique F\'ed\'erale de Lausanne (EPFL), Lausanne, Switzerland}
\author{David Vanderbilt}
\altaffiliation{\WDG}
\affiliation{Department of Physics and Astronomy, Rutgers University, Piscataway, New Jersey 08854-8019, USA}
\author{Ivo Souza}
\altaffiliation{\WDG}
\affiliation{Centro de F\'isica de Materiales, Universidad del Pa\'is Vasco, E-20018 San Sebasti\'an, Spain}
\affiliation{Ikerbasque Foundation, E-48013 Bilbao, Spain}
\author{Arash A. Mostofi}
\altaffiliation{\WDG}
\affiliation{Departments of Materials and Physics, and the Thomas Young Centre for Theory and Simulation of Materials, Imperial College London, London SW7 2AZ, UK}
\author{Jonathan R. Yates}
\altaffiliation{\WDG}
\affiliation{Department of Materials, University of Oxford, Parks Road, Oxford OX1 3PH, UK}

\date{\today}

\begin{abstract}
\Wannier{} is an open-source computer program for calculating maximally-localised Wannier functions (MLWFs) from a set of Bloch states. It is interfaced to many widely used electronic-structure codes thanks to its independence from the basis sets representing these Bloch states.
In the past few years the development of \Wannier{} has transitioned to a community-driven model; this has resulted in a number of new developments that have been recently released in \Wannier{} v3.0. In this article we describe these new functionalities, that include the implementation of new features for wannierisation and disentanglement (symmetry-adapted Wannier functions, selectively-localised Wannier functions, selected columns of the density matrix) and the ability to calculate new properties (shift currents and Berry-curvature dipole, and a new interface to many-body perturbation theory); performance improvements, including parallelisation of the core code; enhancements in functionality (support for spinor-valued Wannier functions, more accurate methods to interpolate quantities in the Brillouin zone); improved usability (improved plotting routines, integration with high-throughput automation frameworks), as well as the implementation of modern software engineering practices (unit testing, continuous integration, and automatic source-code documentation). These new features, capabilities, and code development model aim to further sustain and expand the community uptake and range of applicability, that nowadays spans complex and accurate dielectric, electronic, magnetic, optical, topological and transport properties of materials. 
\end{abstract}

\maketitle

\section{Introduction}

\label{sec:intro}
\Wannier{} is an open-source code for generating Wannier functions (WFs), in particular maximally-localised Wannier functions (MLWFs), and using them to compute advanced materials properties with high efficiency and accuracy. \Wannier{} is a paradigmatic example of interoperable software, achieved by ensuring that all the quantities required as input are entirely independent of the underlying electronic-structure code from which they are obtained. Most of the major and widely used electronic-structure codes have an interface to \Wannier{}, including \QE{}\cite{giannozzi_qe_2017}, ABINIT\cite{gonze-cpc180}, VASP\cite{kresse-prb47,kresse-cms6,kresse-prb54}, Siesta\cite{soler-condmatt14}, Wien2k\cite{blaha-wien2k}, Fleur\cite{blugel-fleur} and Octopus\cite{octopus2015}. As a consequence, once a property is implemented within \Wannier{}, it can be immediately available to users of all codes that interface to it.

Over the last few years, \Wannier{} has undergone a transition from a code developed by a small group of developers to a community code with a much wider developers' base. This has been achieved in two principal ways: (i) hosting the source code and associated development efforts on a public GitHub repository\cite{W90repo}; and (ii) building a community of \Wannier{} developers and facilitating personal interactions between individuals through community workshops, the most recent in 2016. In response, the code has grown significantly, gaining many novel features contributed by this community, as well as numerous fixes.

In this paper, we describe the most important novel contributions to the \Wannier{} code, as embodied in its 3.0 release. The paper is structured as follows: In Sec.~\ref{sec:background} we first summarise the background theory for the computation of MLWFs (additional details can be found in Ref.~\onlinecite{Marzari-RMP2012}), and introduce the notation that will be used throughout the paper. In Sec.~\ref{sec:core} we describe the novel features of \Wannier{}
that are related to the core wannierisation and disentanglement algorithms; these include 
symmetry-adapted WFs, selective localisation of WFs, and
parallelisation using the message-passing interface (MPI). 
In Sec.~\ref{sec:enhancements} we describe new functionality enhancements, including the ability to handle spinor-valued WFs and calculations with non-collinear spin that use ultrasoft pseudopotentials (within \QE{}); improved interpolation of the \mbox{$k$-space} Hamiltonian; a more flexible approach for handling and using initial projections; and the ability to plot WFs in Gaussian cube format on WF-centred grids with non-orthogonal translation vectors. In Sec.~\ref{sec:postproc} we describe new functionalities associated with using MLWFs for computing advanced electronic-structure properties, including the calculation of shift currents, gyrotropic effects and spin Hall conductivities,
as well as parallelisation improvements and the interpolation of bands 
originating from calculations performed with many-body 
perturbation theory (GW). 
In Sec.~\ref{sec:scdm} we describe the selected-columns-of-the-density-matrix (SCDM) method, which enables computation of WFs without the need for explicitly defining initial projections. In Sec.~\ref{sec:aiida} we describe new post-processing tools and codes, and the integration of \Wannier{} with high-throughput automation and workflow management tools (specifically, the AiiDA materials' informatics infrastructure\cite{pizzi-AiiDA}).
In \Secref{sec:goodcodepractices} we describe the modern software engineering practices now adopted in \Wannier{}, that have made it possible to improve the development lifecycle and transform \Wannier{} into a community-driven code. Finally, our conclusions and outlook are presented in Sec.~\ref{sec:conclusions}.

\section{Background}\label{sec:background}
In the independent-particle approximation, the electronic structure of a periodic system is conventionally represented in terms of one-electron Bloch states $\psi_{n\bfk}(\bfr)$, 
which are labelled by a band index $n$ and a crystal momentum $\bfk$ inside the first Brillouin zone (BZ), and which satisfy Bloch's theorem: 
\begin{equation}
\psi_{n\bfk}(\bfr) = u_{n\bfk}(\bfr) e^{i\bfk\cdot\bfr},
\label{eq:bloch_functions}
\end{equation}
where $u_{n\bfk}(\bfr) = u_{n\bfk}(\bfr+\bfR)$ is a periodic function with the same periodicity of the single-particle Hamiltonian, and $\bfR$ is a Bravais lattice vector. 
(For the moment we ignore the spin degrees of freedom and work with spinless
wave functions; spinor wave functions will be treated in Sec.~\ref{sec:spinor}.)
Such a formalism is also commonly applied, via the supercell approximation, to non-periodic systems, typically used to treat point, line and planar defects in crystals, surfaces, amorphous solids, liquids and molecules. 

\subsection{Isolated bands}\label{sec:isolated}

A group of bands is said to be \emph{isolated} if it is separated by energy gaps from all the other lower and higher bands throughout the BZ (this isolated group of bands may still show arbitrary crossing degeneracies and hybridisations within itself). For such isolated set of $J$ bands,
the electronic states can be equivalently represented by a set of $J$ WFs per cell, that are related to the Bloch states via two
unitary transformations (one continuous, one discrete)~\cite{wannier-pr37}:
\begin{equation}
\kwann{n}{\bfR} = V\intBZ{\frac{\diffk}{(2\pi)^3}} e^{-i\bfk\cdot\bfR}\sum_{m=1}^{J} \kpsi{m}{\bfk} \umn,
\label{eq:wannier_transform}
\end{equation}
where $\wannr{n}{\bfR}=w_{n{\bf 0}}(\bfr-\bfR)$ is a periodic (but not necessarily localised) WF labelled by the quantum number $\bfR$ (the counterpart of the quasi-momentum $\bfk$ in the Bloch representation), $V$ is the cell volume and $\matUk$ are unitary matrices that mix Bloch states at a given $\bfk$ and represent the gauge freedom that exists in the definition of the Bloch states and that is inherited by the WFs. 

MLWFs are obtained by choosing \matUk matrices that minimise the sum of the quadratic 
spreads of the WFs about their centres for a reference $\bfR$ (say, $\bfR=\bm{0}$). This sum is given by the spread functional
\begin{equation}
\Omega = \sum_{n=1}^{J} \left[ \Elm{\wann{n}{\bfO}}{\bfr\cdot\bfr}{\wann{n}{\bfO}} - \modulussqd{\Elm{\wann{n}{\bfO}}{\bfr}{\wann{n}{\bfO}}}\right].\label{eq:omega_wannier}
\end{equation}
$\Omega$ may be decomposed into two positive-definite parts\cite{MV_PRB56},
\begin{equation}
\Omega = \omi + \omt,
\label{eq:Omega-decomp}
\end{equation}
where
\begin{equation}
\omi = \sum_n \left[ \Elm{\wann{n}{\bm{0}}}{\bfr\cdot\bfr}{\wann{n}{\bm{0}}} - \sum_{m\bfR} \modulussqd{\Elm{\wann{m}{\bfR}}{\mathbf{r}}{\wann{n}{\bm{0}}}}\right]
\end{equation}
is gauge invariant (i.e., invariant under the action of any unitary $\matUk$ on the Bloch states), and
\begin{equation}
\omt = \sum_{n}\sum_{m\bfR \ne n\bm{0}} \modulussqd{\Elm{\wann{m}{\bfR}}{\mathbf{r}}{\wann{n}{\bm{0}}}}
\end{equation}
is gauge dependent. Therefore, the ``wannierisation" of an isolated manifold of bands, i.e., the transformation of Bloch states into MLWFs, amounts to minimising the gauge-dependent part $\omt$ of the spread functional. 

Crucially, the matrix elements of the position operator between WFs can be expressed in reciprocal space. Under the assumption that the BZ is sampled on a uniform Monkhorst--Pack mesh of \mbox{$k$-points} composed of $N$ points \mbox{($V \int_{\mbox{\tiny{BZ}}}\frac{\diffk}{(2\pi)^3} \rightarrow \frac{1}{N}\sum_\bfk$)}, the gauge-independent and gauge-dependent parts of the spread may be expressed, respectively, as\cite{MV_PRB56}
\begin{equation}\label{eq:omi-k}
\omi = \frac{1}{N} \sum_{\bfk,\bfb} w_b \left[ J - \sum_{mn} \modulussqd{\mmn}\right]
\end{equation}
and
\begin{equation}\label{eq:omt-k}
\begin{aligned}
\omt & = \frac{1}{N} \sum_{\bfk,\bfb} w_b \sum_{m\ne n} \modulussqd{\mmn} \\
& \qquad + \frac{1}{N} \sum_{\bfk,\bfb} w_b \sum_{n} (-\:\text{Im}\ln \mnn - \bfb\cdot\bar{\bfr}_n)^2,
\end{aligned}
\end{equation}
where $\bfb$ are the vectors connecting a \mbox{$k$-point} to its neighbours, $w_b$ are weights associated
with the finite-difference representation of $\nabla_{\bfk}$ for a given geometry, the matrix of overlaps $\matMkb$ is defined by
\begin{equation}
\mmn = \inprod{u_{m\bfk}}{u_{n,\bfk+\bfb}},\label{eq:mmn}
\end{equation} 
and the centres of the WFs are given by
\begin{equation}
\bar{\bfr}_n \equiv \Elm{\wann{n}{\bm{0}}}{\mathbf{r}}{\wann{n}{\bm{0}}} = -\frac{1}{N}\sum_{\bfk,\bfb} w_b \bfb\: \text{Im}\ln \mnn.
\end{equation}
Minimisation of the spread functional is achieved by considering infinitesimal gauge transformations \mbox{$\umn = \delta_{mn} + \fulld W_{mn\bfk}$}, where $\fulld W$ is anti-Hermitian \mbox{($\fulld W^\dag = - \fulld W$)}. The gradient of the spread functional with respect to such variations is given by
\begin{equation}
    \matcalGk \equiv \totald{\Omega}{W_{mn\bfk}} = 4 \sum_{\bfb} w_{b}\left(\mathcal{A}[R_{mn}^{(\bfk,\bfb)}]-\mathcal{S}[T_{mn}^{(\bfk,\bfb)}]\right),
    \label{eq:gradient-omega}
\end{equation}
where $\mathcal{A}$ and $\mathcal{S}$ are the super-operators $\mathcal{A}[B]=(B-B^\dag)/2$ and $\mathcal{S}[B]=(B+B^\dag)/2i$, respectively, and
\begin{align}
   R_{mn}^{(\mathbf{k,b})} &= \mmn M_{nn}^{(\mathbf{k,b})*}, \label{eq:Rmn} \\
   T_{mn}^{(\mathbf{k,b})} &= \frac{\mmn}{\mnn} q_{n}^{(\mathbf{k,b})}, \label{eq:Tmn} \\
   q_{n}^{(\mathbf{k,b})} &= \text{Im} \ln \mnn + \bfb\cdot\bar{\bfr}_n.
\end{align}
For the full derivation of Eq.~(\ref{eq:gradient-omega}) we refer to Ref.~[\onlinecite{MV_PRB56}]. 
This gradient is then used to generate a search direction $\matcalDk$ for an iterative steepest-descent or conjugate-gradient minimisation of the spread~\cite{Mostofi_CPC}: at each iteration the unitary matrices are updated according to
\begin{equation}\label{eq:U-update}
    \matUk \rightarrow \matUk\:\text{exp}[\alpha \matcalDk],
\end{equation}
where $\alpha$ is a coefficient that can either be set to a fixed value or determined at each iteration via a simple polynomial line-search, and the matrix exponential is computed in the diagonal representation of $\matcalDk$ and then transformed back in the original representation. Once the unitary matrices have been updated, the updated set of $\matMkb$ matrices is calculated according to
\begin{equation}
    \label{eq:updated-Mmn}
    \matMkb = \matUk^{\dagger} M^{(0)(\bfk,\bfb)} U^{\phantom{\dagger}}_{\bfk+\bfb},
\end{equation}
where 
\begin{equation}
    \label{eq:def-Mmn}
    M_{mn}^{(0)(\mathbf{k,b})} = \inprod{u^{(0)}_{m\bfk}}{u^{(0)}_{n,\bfk+\bfb}}
\end{equation}
is the set of initial $\matMkb$ matrices, computed once and for all, at the start of the calculation, from the original set of reference Bloch orbitals $\ket{u^{(0)}_{n\bfk}}$. 

\subsection{Entangled bands}\label{sec:entangled}
It is often the case that the bands of interest are not separated from other bands in the Brillouin zone by energy gaps and are overlapping and hybridising with other bands that extend  beyond the energy range of interest. In such cases, we refer to the bands as being \emph{entangled}. 

The difficulty in constructing MLWFs for entangled bands arises from the fact that, within a given energy window, the number of bands $\mathcal{J}_{\bfk}$ at each $k$-point $\bfk$ in the BZ is not a constant and is, in general, different from the target number $J$ of WFs: $\mathcal{J}_{\bfk} \ge J$. Even making the energy window $k$-dependent would see discontinuous inclusion and exclusion of bands as the BZ is traversed. The treatment of entangled bands requires thus a more complex approach that is typically a two-step process. In the first step, a $J$-dimensional manifold of Bloch states is selected at each $k$-point, chosen to be as smooth as possible as a function of $\bfk$. In the second step, the gauge freedom associated with the selected manifold is used to obtain MLWFs, just as described in Sec.~\ref{sec:isolated} for the case of an isolated set of bands.

Focusing on the first step, an orthonormal basis for the $J$-dimensional subspace $\mathcal{S}_\bfk$ at each $\bfk$ can be obtained by performing a semi-unitary transformation on the $\mathcal{J}_{\bfk}$ states at $\bfk$,
\begin{equation}
    \ket{\widetilde{\psi}_{n\bfk}}=\sum_{m=1}^{\mathcal{J}_{\bfk}} \ket{\psi_{m\bfk}} \vmn,
\end{equation}
where $\matVk$ is a rectangular matrix of dimension $\mathcal{J}_{\bfk}\times J$ that is semi-unitary in the sense that $\matVk^{\dagger}\matVk^{\phantom{\dagger}}=\mathbf{1}$.

To select the smoothest possible manifold, a measure of the intrinsic smoothness of the chosen subspace is needed. It turns out that such a measure is given precisely by what was the gauge-invariant part $\omi$ of the spread functional for isolated bands.\cite{SMV_PRB65} Indeed, \Eqref{eq:omi-k} can be expressed as
\begin{equation}
    \omi=\frac{1}{N} \sum_{\bfk,\bfb} w_{b} \Trace [{P}_{\bfk}{Q}_{\bfk+\bfb}],
\end{equation}
where ${P}_{\bfk}=\sum_{n=1}^{J}\Ketbra{\widetilde{u}_{n\bfk}}{\widetilde{u}_{n\bfk}}$ is the projection operator onto $\mathcal{S}_\bfk$, ${Q}_\bfk={\mathbf{1}}-{P}_\bfk$ is its Hilbert-space complement, and ``$\Trace$" represents the trace over the entire Hilbert space. $\Trace [{P}_{\bfk}{Q}_{\bfk+\bfb}]$ measures the mismatch between the subspaces $\mathcal{S}_\bfk$ and $\mathcal{S}_{\bfk+\bfb}$, vanishing if they overlap identically. Hence $\omi$ measures the average mismatch of the local subspace $\mathcal{S}_\bfk$ across the BZ, so that an optimally-smooth subspace can be
selected by minimising $\omi$. Doing this with orthonormality constraints on the Bloch-like states is equivalent to solving self-consistently the set of coupled eigenvalue equations\cite{SMV_PRB65}
\begin{equation}
    \left[\sum_{\bfb} w_{b} {P}_{\bfk+\bfb}\right] \ket{\widetilde{u}_{n\bfk}} = \lambda_{n\bfk}\ket{\widetilde{u}_{n\bfk}}.
\end{equation}
The solution can be achieved via an iterative procedure, whereby at the $i^{\mathrm{th}}$ iteration the algorithm traverses the entire set of \mbox{$k$-points}, selecting at each one the $J$-dimensional subspace $\mathcal{S}^{(i)}_{\bfk}$ that has the smallest mismatch with the subspaces $\mathcal{S}^{(i-1)}_{\bfk+\bfb}$ at the neighbouring \mbox{$k$-points} obtained in the previous iteration. This amounts to solving 
\begin{equation}\label{eq:disentangle1}
    \left[\sum_{\bfb} w_{b} {P}_{\bfk+\bfb}^{(i-1)}\right] \ket{\widetilde{u}_{n\bfk}^{(i)}} = \lambda_{n\bfk}^{(i)}\ket{\widetilde{u}_{n\bfk}^{(i)}},
\end{equation}
and selecting the $J$ eigenvectors with the largest eigenvalues\cite{SMV_PRB65}. Self-consistency is reached when $\mathcal{S}^{(i)}_{\bfk} = \mathcal{S}^{(i-1)}_{\bfk}$ (to within a user-defined threshold \texttt{dis\_conv\_tol}) at all the \mbox{$k$-points}. To make the algorithm more robust, the projector appearing on the left-hand-side of \Eqref{eq:disentangle1} is replaced with $[{P}_{\bfk+\bfb}^{(i)}]_{\mathrm{in}}$, given by
\begin{equation}
    [{P}_{\bfk+\bfb}^{(i)}]_{\mathrm{in}} = \beta {P}_{\bfk+\bfb}^{(i-1)} + (1-\beta)[{P}_{\bfk+\bfb}^{(i-1)}]_{\mathrm{in}},
\end{equation}
which is a linear mixture of the projector that was used as input for the previous iteration and the projector defined by the output of the previous iteration. The parameter $0 < \beta \le 1$ determines the degree of mixing, and is typically set to $\beta=0.5$; setting $\beta=1$ reverts precisely to \Eqref{eq:disentangle1}, while smaller and smaller values of $\beta$ make convergence smoother (and thus more robust) but also slower.

In practice, \Eqref{eq:disentangle1} is solved by diagonalising the Hermitian operator appearing on the left-hand-side in the basis of the original $\mathcal{J}_{\bfk}$ Bloch states: 
\begin{equation}
    Z_{mn\bfk}^{(i)} = \elm{u_{m\bfk}^{(0)}}{\sum_{\bfb} w_{b} [{P}_{\bfk+\bfb}^{(i)}]_{\mathrm{in}}}{u_{n\bfk}^{(0)}}.
    \label{eq:z_matrix}
\end{equation}

Once the optimal subspace has been selected, the wannierisation procedure described in Sec.~\ref{sec:isolated} is carried out to 
minimise the gauge-dependent part $\omt$ of the spread functional within that optimal subspace.

\subsection{Initial projections}

In principle, the overlap matrix elements $\mmn$ are the only quantities required to compute and minimise the spread functional, and generate MLWFs for either isolated or entangled bands. In practice, this is generally true when dealing with an isolated set of bands, but in the case of entangled bands a good initial guess for the subspaces $\mathcal{S}_\bfk$ alleviates problems associated with falling into local minima of $\omi$, and/or obtaining MLWFs that cannot be chosen to be real-valued (in the case of spinless WFs). Even in the case of an isolated set of bands, a good initial guess for the WFs, whilst not usually critical, often results in faster convergence of the spread to the global minimum. (It is important to note that both for isolated and for entangled bands multiple solutions to the wannierisation or disentanglement can exist, as discussed later.) 

A simple and effective procedure for selecting an initial gauge (in the case of isolated bands) or an initial subspace and initial gauge (in the case of entangled bands) is to project a set of $J$ trial orbitals $g_n(\bfr)$ localised in real space onto the space spanned by the set of original Bloch states at each $\bfk$:
\begin{equation}
    \ket{\phi_{n\bfk}} = \sum_{m=1}^{J\:\mathrm{or}\:\mathcal{J}_{\bfk}} \ket{\psi_{m\bfk}}\inprod{\psi_{m\bfk}}{g_n},
\end{equation}
where the sum runs up to either $J$ or $\mathcal{J}_{\bfk}$, depending on whether the bands are isolated or entangled, respectively, and the inner product $\amn=\inprod{\psi_{m\bfk}}{g_n}$ is over all the Born--von Karman supercell. (In practice, the fact that the $g_n$ are localised greatly simplifies this calculation.) The  matrices $\matAk$ are square $(J\times J)$ or rectangular $(\mathcal{J}_{k}\times J)$ in the case of isolated or entangled bands, respectively. The resulting orbitals are then orthonormalised via a L\"{o}wdin transformation~\cite{lowdin1950}:
\begin{eqnarray}
    \ket{\widetilde{\psi}_{n\bfk}} &=& \sum_{m=1}^{J} \ket{\phi_{m\bfk}} \smn^{-\nicefrac{1}{2}}\\ 
    &=& \sum_{m=1}^{J\:\mathrm{or}\:\mathcal{J}_{\bfk}} \ket{\psi_{m\bfk}} (\matAk\matSk^{-\nicefrac{1}{2}})_{mn},
    \label{eq:initial-U}
\end{eqnarray}
where
$\smn={\inprod{\phi_{m\bfk}}{\phi_{n\bfk}}}=(\matAk^{\dagger}\matAk^{\phantom{\dagger}})_{mn}$, and $\matAk^{\phantom{}}\matSk^{-\nicefrac{1}{2}}$ 
is a unitary or semi-unitary matrix. In the case of entangled bands, once an optimally-smooth subspace has been obtained as described in Sec.~\ref{sec:entangled}, the same trial orbitals $g_n(\bfr)$ can be used to initialise the wannierisation procedure of Sec.~\ref{sec:isolated}. In practice, the matrices $\matAk$ are computed once and for all at the start of the calculation, together with the overlap matrices $\matMkb$.
These two operations need to be performed within the context of the electronic-structure code and basis set adopted; afterwards, all the operations of \Wannier{} rely only on $\matAk$ and $\matMkb$ and not on the specific representation of $\psi_{m\bfk}$ (e.g., plane waves, linearised augmented plane waves, localised basis sets, real-space grids, \ldots).

\section{New features for wannierisation and disentanglement}\label{sec:core}
In this section we provide an overview of the new features associated with the core wannierisation and disentanglement algorithms in \Wannier{}, namely the ability to generate WFs of specific symmetry; selectively localise a subset of the WFs and/or constrain their centres to specific sites; and perform wannierisation and disentanglement more efficiently through parallelisation.

\subsection{Symmetry-adapted Wannier functions}\label{subsubsec:sawfs}
In periodic systems, 
atoms are usually found at sites $\bfq$ whose site-symmetry group $G_q$ is a subgroup of the full point group $F$ of the crystal\cite{evarestov2012site} (the symmetry operations in the group $G_q$ are those that leave $\bfq$ fixed). 
The set of points $\bfq_a$ that are symmetry-equivalent sites to $\bfq$ is called an {\it orbit}\cite{Int_tab_cryst}. 
These are all the points in the unit cell that can be generated from $\bfq$ by applying the symmetry operations in $G$ that do not leave $\bfq$ fixed.
If $\bfq_a$ is a high-symmetry site then its Wyckoff position has a single orbit\cite{Int_tab_cryst}; for low-symmetry sites different orbits correspond to the same Wyckoff position.
The number of points in the orbit(s) is the multiplicity $n_{q_a}$ of the Wyckoff position.  
\MLWFs{}, however, are not bound to reside on such high-symmetry sites, and they do not necessarily possess the site symmetries of the crystal \cite{SMV_PRB65, Sakuma_PRB87, Thygesen2005}. When using \MLWFs{} as a local orbital basis set in methods such as first-principles tight binding, DFT+U and DFT plus dynamical-mean-field theory (DMFT), which deal with beyond-DFT correlations in a local subspace such as that spanned by $3d$ orbitals (for transition metals or transition-metal oxides)
or $4f$ orbitals (for rare-earth or actinide intermetallics), it is often desirable to ensure that the WFs basis possesses the local site symmetries. 

Sakuma\cite{Sakuma_PRB87} has shown that such {\it symmetry-adapted Wannier functions} (SAWFs) can be constructed by introducing additional constraints on the unitary matrices $\matUk$ of Eq.~\eqref{eq:wannier_transform} during the minimisation of the spread. SAWFs, therefore, can be fully integrated within the original maximal-localisation procedure. The SAWF approach gives the user a certain degree of control over the symmetry and centres of the Wannier functions at the expense of some localisation since the final total spread of the resulting SAWFs can only be equal to, or most often larger than, that of the corresponding \MLWFs{} with no constraints (note that in principle some SAWFs can have a smaller individual spread than any \MLWFs{}). 

A set of SAWFs 
\begin{equation}
\{w_{i}^{(\varrho)}(\bfr - \bfq_ a  )=w_{i a  }^{(\varrho)}(\bfr),\quad i=1,\dots,n_\varrho\}
\end{equation}
can be specified by one representative point $\bfq_a$ of the orbit (in the home unit cell $\bfR=\bfO$), and by the irreducible representation ({\it irrep}) $\varrho$ of the corresponding site-symmetry group $G_{a}$ (the dimension of the {\it irrep} being $n_\varrho$). For instance, in simple fcc crystals, like copper (space group $Fm{-}3m$), the Cu atom is placed at the origin of the unit cell $\bfq=(0,0,0)$ (Wyckoff letter $a$ with multiplicity 1, i.e., only one point in the orbit of $(0,0,0)$ in the unit cell, due to the fact that $Fm{-}3m$ is symmorphic\cite{Int_tab_cryst}), whose site-symmetry group is $m{-}3m$ (also referred to as $O_h$). One of the {\it irreps} of $O_h$ is that with character $T_{2g}$, which is 3-dimensional.

To find symmetry-adapted Wannier functions, one needs to specify the unitary transformations $U^{(\varrho)}_{m i a\bfk}$ of the Bloch states, defined by
\begin{eqnarray}
w_{i a}^{(\varrho)}(\bfr-\bfR) &=& \frac{1}{N} \sum_{\bfk} e^{-i\bfk\cdot\bfR}\sum_{m=1}^J  \psi_{m\bfk}(\bfr) U_{m i a\bfk}^{(\varrho)} \nonumber \\ 
&=& \frac{1}{N}\sum_{\bfk} e^{-i\bfk\cdot\bfR} \psi_{i a  \bfk}^{(\varrho)}(\bfr).
\label{eq:bloch_lincomb}
\end{eqnarray}
Therefore, the goal is to construct basis functions of the {\it irrep} $\varrho$, $\{\psi_{i a  \bfk}^{(\varrho)}(\bfr)\}$ 
, from a linear combination of the $J$ eigenstates $\psi_{n\bfk}(\bfr)$  of the Hamiltonian $H$. 
Since $H$ is invariant under the full space-group $G$, the representation of a given symmetry operation \mbox{$g=(\mathcal{R}\vert \mathbf{t}) \in G$} (where $\mathcal{R}$ and $\mathbf{t}$ are the rotation and fractional-translation parts of the symmetry operation, respectively) in the $\{\psi_{n\bfk}(\bfr)\}$ basis must be a $J\times J$ unitary matrix\cite{evarestov2012site} $\widetilde{d}_\bfk(g)$, \ie{} $\widetilde{d}_\bfk(g)$ represents how the $J$ Bloch states are transformed by the symmetry operation $g$:
\begin{equation}
    {g}\,\psi_{n\bfk}(\bfr) = \sum_{m=1}^{J}\psi_{m\mathcal{R}\bfk}(\bfr)\widetilde{d}_{mn\bfk}(g), \quad {g} \in G.
    \label{eq:g_on_bf}
\end{equation}
When single-particle eigenfunctions $\psi_{n\bfk}(\bfr)$ are used, as in this case, the matrix elements $\widetilde{d}_{\bfk}(g)$ can be computed directly as  
\begin{equation}
    \widetilde{d}_{mn\bfk}(g) = \int\!\!\diffr \, \psi_{m\mathcal{R}\bfk}^\ast(\bfr) \psi_{n\bfk}\left(g^{-1}\bfr\right).
\label{eq:d_bands}\end{equation}
As for the overlap matrices $\matMkb$, the resulting $\widetilde{d}_{\bfk}(g)$ are also basis-set independent. Moreover, once computed (using the original gauge) they remain fixed during the calculation. For instance, in a plane-wave code the integrals in Eq.~\eqref{eq:d_bands} can be easily computed in reciprocal space.

On the other hand, it can be shown that the SAWFs 
transform under the action of $g \in G$ with a different matrix $d^{(\varrho)}_\bfk(g)$ \cite{Sakuma_PRB87,evarestov2012site}, which in turn defines how the $\psi_{i a  \bfk}^{(\varrho)}(\bfr)$ transform under the action of $g\in G$: 
\begin{eqnarray}
    {g} \psi_{i a  \bfk}^{(\varrho)}(\bfr) &=& \sum_{i' a' \varrho'}  \psi_{i' a' \mathcal{R}\bfk}^{(\varrho')}(\bfr)D_{i' a' ,i a   \bfk}^{(\varrho',\varrho)}(g),
        \label{eq:g_on_sabf}
\end{eqnarray}
where 
\begin{equation}
    D_{i' a' ,i a  \bfk}^{(\varrho',\varrho)}(g)=\delta_{\varrho',\varrho}e^{-i\mathcal{R}\bfk\cdot\bfR_{ a'  a  }}d_{i'i\bfk}^{(\varrho)}(g)\label{eq:d_wann}
\end{equation}
and
\begin{equation}
\bfR_{a'a}=g\bfq_{a}   - \bfq_{a'}
\end{equation}
is a lattice translation vector.  It is worth to mention that ${ a'  }$ in Eq.~\eqref{eq:g_on_sabf} is uniquely defined by specifying the symmetry operations $g\in G$; see Ref.~\onlinecite{evarestov2012site,Sakuma_PRB87} for details.

$D_{\bfk}(g)$ is block-diagonal in the $\varrho$ index.  For a given set of Wyckoff positions, the number of blocks is given by the sum of the number of all \textit{irreps} considered (if non-equivalent Wyckoff positions ($ \bfq_b  \neq  \bfq_a   $) are present then $D_{\bfk}(g)$ contains also blocks corresponding to these positions). Each block contains $n_{q_a}^2$ sub-matrices of dimension $n_\varrho\times n_\varrho$. Therefore, if only one Wyckoff position is given with multiplicity $n_{q_a}$, then there are $J=n_{q_a} n_\varrho$ energy bands in the representation given by the $\{\psi_{i a  \bfk}^{(\varrho)}(\bfr) \}$  (see Ref.~\onlinecite{Sakuma_PRB87} for full details).

To compute the $D_{\bfk}(g)$ matrices one needs to specify the centre $\bfq_a$ and the symmetry of the initial functions (e.g., $s, p,$ and $d$). Then, for each symmetry operation $g_{q_a}$ in the site-symmetry group $G_{q_a}$ one needs to calculate the matrix representation of the rotational part expressed in the basis of these functions.

From Eqs.~(\ref{eq:bloch_lincomb}), (\ref{eq:g_on_bf}) and (\ref{eq:g_on_sabf}) one can show that the following relationship exists between $\matUk$ and $U_{\mathcal{R}\bfk}$
\begin{equation}
    U_{\mathcal{R}\bfk}D_{\bfk}(g) = \widetilde{d}_{\bfk}(g)\matUk.
    \label{eq:sawfs-constraint}
\end{equation}
Let $g_\bfk$ now be the symmetry operations that leave a given $\bfk$ unchanged. Then, Eq.~\eqref{eq:sawfs-constraint} gives the condition that $\matUk$ must satisfy in this case:
\begin{equation}
    \matUk D_{\bfk}(g_\bfk) = \widetilde{d}_{\bfk}(g_\bfk)\matUk, \quad g_\bfk \in G_\bfk.
    \label{eq:sawfs-constraint_littlek}
\end{equation}

The initial unitary matrix $\matUk$ ($\bfk \in$ IBZ) must satisfy the constraints in Eq.~(\ref{eq:sawfs-constraint_littlek}); this can be done in an iterative fashion, as discussed in Ref.~\onlinecite{Sakuma_PRB87}. 
In practice, the Wannier functions are generated from a limited subspace spanned by a finite number of states inside a target ``energy window'', but this does not guarantee that a $\matUk$ can be constructed for any desired {\it irrep}. In fact, if a given {\it irrep} is not compatible with the symmetry of the states within the energy window, Eq.~\eqref{eq:sawfs-constraint_littlek} cannot be fulfilled. 
 
For an isolated set of bands, the minimisation of $\omt$ with the constraints defined in Eq.~(\ref{eq:sawfs-constraint}) requires 
the gradient $\mathcal{G}\tinysup{sym}_{\bfk}$ of the total spread $\Omega$ with respect to a symmetry-adapted gauge variation to generate a search direction $\mathcal{D}\tinysup{sym}_{\bfk}$. The equation for the symmetry-adapted gradient reads
\begin{equation}
    \mathcal{G}\tinysup{sym}_{\bfk} = \frac{1}{n_\bfk} \sum_{g=(\mathcal{R}\vert \mathbf{t})\in G} D_{\bfk}(g)\mathcal{G}_{\mathcal{R}\bfk}D_{\bfk}^{\dag}(g),
    \label{eq:sawfs-gradient}
\end{equation}
where $\mathcal{G}_{\bfk}$ is the original gradient given in Eq.~(\ref{eq:gradient-omega}), and $n_\bfk$ is the number of symmetry operations in $G$ that leave $\bfk$ fixed. 

The procedure described above for isolated bands has to be modified only slightly for the case of entangled bands. The main difference with respect to the unconstrained case of Sec.~\ref{sec:background} is that the $J$  eigenvectors of the $J$ largest eigenvalues of the $Z_\bfk$ matrix in Eq.~(\ref{eq:z_matrix}) do not necessarily span the same subspace spanned by the desired symmetry-adapted Wannier functions. 
Since direct minimisation is not bound to give symmetry-adapted WFs, Sakuma\cite{Sakuma_PRB87} has proposed an alternative steepest-descent approach to construct the optimal unitary matrix from a set of $\mathcal{J}_{\bfk} \geq J$ Bloch wavefunctions that also fulfil symmetry constraints. 
Once this step is completed and optimal symmetry-adapted Bloch functions $\widetilde{\psi}_{ia\bfk}^{(\varrho)}(\bfr)$ have been computed, the algorithm proceeds as in the isolated case where one seeks the $\matUk$ that minimise $\omt$ and give the symmetry-adapted Bloch functions in terms of $\widetilde{\psi}_{ia\bfk}^{(\varrho)}(\bfr)$ as in Eq.~\eqref{eq:bloch_lincomb}.

\subsection{Selectively-localised Wannier functions and constrained Wannier centres}\label{subsubsec:slwfs}
Wang \textit{et al.} have proposed an alternative method\cite{Marianetti_PRB_90} to the symmetry-adapted Wannier functions described in Section~\ref{subsubsec:sawfs}. Their method permits the selective localisation of a \textit{subset} of the Wannier functions, which may optionally be constrained to have specified centres. Whilst this method does not enforce or guarantee symmetry constraints, it has been observed in the cases that have been studied\cite{Marianetti_PRB_90} that Wannier functions whose centres are constrained to a specific site typically possess the corresponding site symmetries.

For an isolated set of $J$ bands, selective localisation of a subset of $J'\leq J$ Wannier functions is accomplished by minimising the total spread $\Omega$ with respect to  only $J'\times J'$ degrees of freedom in the unitary matrix $\matUk$. The spread functional to minimise is then given by
\begin{equation}
    \Omega'=\sum_{n=1}^{J'\leq J}\left[\elm{\wann{n}{\bf0}}{r^2}{\wann{n}{\bf0}} - \modulussqd{\elm{\wann{n}{\bf0}}{\bfr}{\wann{n}{\bf0}}} \right],
    \label{eq:omega_wannier_sel}
\end{equation}
which reduces to the original spread functional $\Omega$ of Eq.~(\ref{eq:omega_wannier}) for $J'=J$. When $J'<J$, it is no longer possible to cast the functional $\Omega'$ as a sum of a gauge-independent term $\omi$ and gauge-dependent one $\omt$, as done in Eq.~(\ref{eq:Omega-decomp}) for $\Omega$. Nevertheless, the minimisation can be carried out with methods very similar to those described in Section~\ref{sec:background}. In fact, for $J'<J$, $\Omega'$ can be written as the sum of two gauge-dependent terms, $\Omega'= \omiod + \omd$, where $\omiod$ is formally given by the sum of $\omi$ and the off-diagonal term $(m\neq n), \quad m,n\leq J' < J$ of $\omt$, and $\omd$ by the diagonal term $(m=n)$ of $\omt$. If one adopts the usual discrete representation on a uniform Monkhorst--Pack grid of \mbox{$k$-points}, $\omiod$ and $\omd$ are given by\cite{Marianetti_PRB_90}
\begin{equation}
\omiod = \frac{1}{N}\sum_{\bfk,\bfb} w_b \left[ J' - \sum_{n}^{J'< J}\modulussqd{\mnn}\right] 
\end{equation}
and 
\begin{equation}
\omd = \frac{1}{N}\sum_{n=1}^{J' < J}\sum_{\bfb,\bfk}w_b \left(\:\mrm{Im}\ln\mnn + \bfb\cdot\bar{\bfr}_n \right)^2.
\end{equation}
With this new spread functional, we can mimic the procedure used to obtain a set of \MLWFs{}, and derive the gradient $\mathcal{G}_{\bfk}^{\prime}$ of $\Omega'$ which gives the search direction to be used in the minimisation. The matrix elements of $\mathcal{G}_{\bfk}^{\prime}$ read
\begin{equation}
   \mathcal{G}^{\prime}_{mn\bfk} = \begin{cases} \mathcal{G}_{mn\bfk} & \scalebox{0.6}{$m\leq J',n\leq J'$,} \\
   { -\:2\sum_{\bfb} w_b \left[R_{nm}^{(\bfk,\bfb)\ast} -i T_{nm}^{(\bfk,\bfb)\ast}\right]} & \scalebox{0.6}{$m\leq J', J' < n \leq J$,}\\ 
   { 2\sum_{\bfb} w_b \left[R_{mn}^{(\bfk,\bfb)} +i T_{mn}^{(\bfk,\bfb)}\right]} & \scalebox{0.6}{$J' < m \leq J, n\leq J'$,}\\ 
   {0} & \scalebox{0.6}{$ J' < m \leq J, J' < n \leq J$,}\end{cases}
\end{equation}
where $\mathcal{G}_{mn\bfk}$ are the matrix elements of the original gradient in Eq.~(\ref{eq:gradient-omega}) (see also  Ref.~\onlinecite{MV_PRB56}), and  $R_{mn}^{(\bfk,\bfb)}$ and  $T_{mn}^{(\bfk,\bfb)}$ are given by Eq.~(\ref{eq:Rmn}) and Eq.~(\ref{eq:Tmn}), respectively. As a result of the minimisation, we obtain a set of $J'$ maximally-localised Wannier functions, known as \emph{selectively-localised Wannier functions} (\SLWFs), whose spreads are in general smaller than the corresponding \MLWFs. Naturally, the remaining $J-J'$ functions will be more delocalised than their \MLWF{} counterparts, as they are not optimised, and the overall sum of spreads will be larger (or in the best case scenario equal).

The centres of the \SLWFs{} may be constrained by adding a quadratic penalty function to the spread functional $\Omega'$, defining a new functional given by
\begin{eqnarray}
    \label{eq:omega_lambda}
    \Omega'_{\lambda} &=& \sum_{n=1}^{J'\leq J}[\elm{\wann{n}{\bf0}}{r^2}{\wann{n}{\bf0}} - \modulussqd{\elm{\wann{n}{\bf0}}{\bfr}{\wann{n}{\bf0}}} \\ \nonumber 
    & & + \lambda (\bar{\bfr}_n - \bfx_n)^2],
\end{eqnarray}
where $\lambda$ is a Lagrange multiplier and $\bfx_n$ is the desired centre for the $n^{\rm th}$ WF. 
The procedure outlined above for  minimising $\Omega'$ can be also adapted to deal 
with $\Omega'_{\lambda}$ (see Ref.~\onlinecite{Marianetti_PRB_90} for details), and minimising $\Omega'_{\lambda}$ results in selectively-localised Wannier functions subject to the constraint of fixed centres (SLWF+C). As noted above, it is observed that WFs derived using the SLWF+C approach naturally possess site symmetries, and their individual spreads are usually smaller than the corresponding spreads of \MLWFs{}, although the total spread, combination of the $J'$ selectively optimised WFs and the $J-J'$ unoptimised functions, is larger than the total spread of the \MLWFs{} (see, for instance last column in Tab.~\ref{fig:slwf_sawf}). 

In the case of entangled bands, the SLWF(+C) method implicitly assumes that a subspace selection has been performed, \ie{}, that a smooth $J$-dimensional manifold exists. Since for the $\Omega'$ and $\Omega'_{\lambda}$ functionals it is not possible to define an $\Omega_I$ that measures the intrinsic smoothness of the underlying manifold, the additional constraints in Eq.~\eqref{eq:omega_wannier_sel} and Eq.~\eqref{eq:omega_lambda} can only be imposed during the wannierisation step. This means that SLWF(+C) can be seamlessly coupled with the disentanglement procedure, with no further additions to the original procedure of Sec.~\ref{sec:entangled}.

\subsection*{SAWF and SLWF+C in GaAs}
As an example of the capabilities of the SAWF and SLWF+C approaches, we show how to construct atom-centred WFs that possess the local site symmetries in gallium arsenide (GaAs). In particular, we discuss how to obtain one WF from the four valence bands of GaAs that is centred on the As atom and that transforms like the identity under the symmetry operations in $T_d$, the site-symmetry group of the As site (for completeness, we also show one \MLWF{} and one SLWF without constraints).
Since we only deal with the four valence bands of GaAs---an isolated manifold---no prior subspace selection is required for the wannierisation.
All calculations were carried out with the plane-wave DFT code Quantum ESPRESSO\cite{giannozzi_qe_2017},
employing PAW pseudopotentials\cite{PAW_PRB50,Kresse_PRB59} from the pslibrary (v1.0)\cite{DALCORSO2014337}. For the exchange-correlation functional we use the Perdew--Burke--Ernzerhof approximation\cite{PBE_PRL77}. The energy cut-off for the plane-waves basis is set to 35.0~Ry, and a $4\times4\times4$ uniform grid is used to sample the Brillouin zone. The lattice parameter is set to the experimental value (5.65 \AA). The overlap matrices $\mmn$ in Eq.~\eqref{eq:mmn}, the projection matrices $\amn$ in Eq.~\eqref{eq:initial-U} and both $\widetilde{d}_{\bfk}(g)$ in Eq.~\eqref{eq:d_bands} and $D_\bfk(g)$ in Eq.~\eqref{eq:d_wann} have been computed with the \pwtowan{} interface.

GaAs is a III-V semiconductor that crystallises in the fcc cubic structure, with a two-atom basis: the Ga cation and the As anion (space group $F{-}43m$); in our example the Ga atom is placed at the origin of the unit cell, whose Wyckoff letter is $a$ and 
site-symmetry group is ${-}43m$, also known as $T_d$. 
The As atom is placed at (\nicefrac{1}{4},\nicefrac{1}{4},\nicefrac{1}{4}), whose Wyckoff letter is $c$ and site-symmetry group is also $T_d$.

\begin{figure}[htp]
    \centering
    \subfloat[MLWF]{\includegraphics[width=3.5cm,trim={100pt 100pt 100pt 50pt},clip]{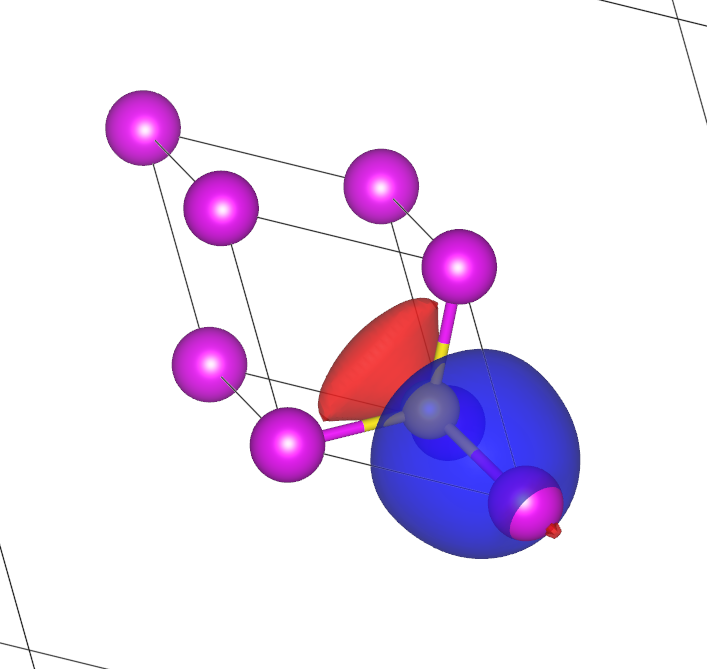}}
    \subfloat[SAWF]{\includegraphics[width=3.5cm,trim={100pt 100pt 100pt 50pt},clip]{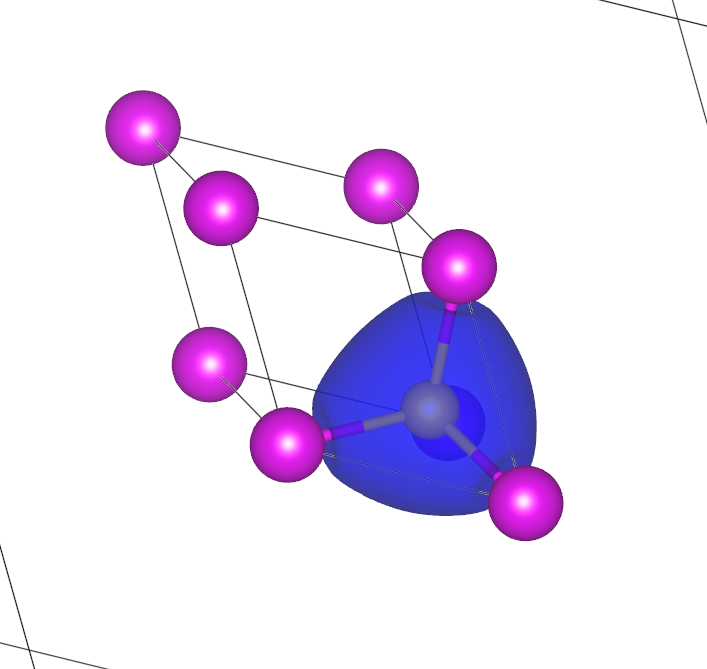}}\\
    \subfloat[SLWF]{\includegraphics[width=3.5cm,trim={100pt 100pt 100pt 50pt},clip]{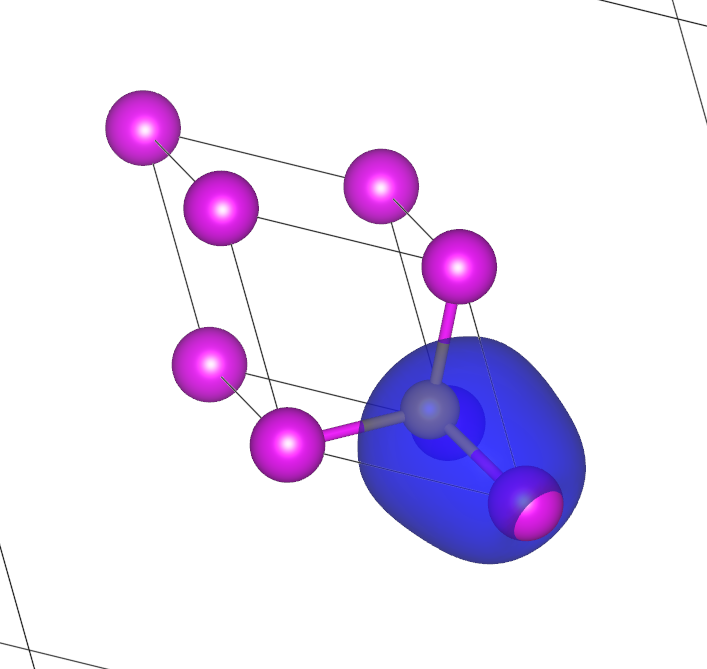}}
    \subfloat[SLWF+C]{\includegraphics[width=3.5cm,trim={100pt 100pt 100pt 50pt},clip]{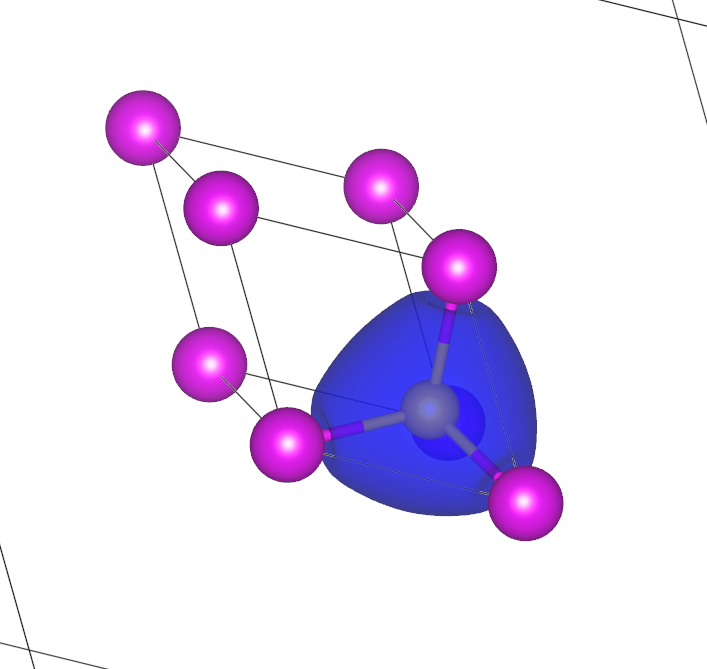}}
    \\
    \vspace{1cm}
\centering
\begin{tabular}{@{} lccc @{}}\toprule[1.5pt]
Method & $\overbar{\bfr}$ (\AA) & $\braket{r^2}-\overbar{r}^2$ (\AA$^2$) & $\Omega$ (\AA$^2$) \\\midrule[1pt]
MLWF & $(-0.857,0.857,0.857)$ & 1.780 & 7.1204\\
SAWF & $(-1.4129,1.4129,1.4129)$ & 1.637 & 10.1365 \\
SLWF & $(-0.89, 0.89, 0.89)$ & 1.424 & 9.8065\\
SLWF+C & $(-1.4129,1.4129,1.4129)$ & 1.634 & 7.8673 \\\bottomrule[1.5pt]
\end{tabular}
\caption{Top (figure): comparison of Wannier functions resulting from different minimisation schemes in gallium arsenide (larger pink spheres are Ga cation atoms and yellow spheres are As anions): a) MLWF; b) SAWF; c) SLWF and d) SLWF+C. For MLWF, SLWF and SLWF+C, four $s$-type orbitals centred at the midpoints of the four Ga--As bonds ((\nicefrac{1}{8},\nicefrac{1}{8},\nicefrac{1}{8}),(\nicefrac{1}{8},\nicefrac{1}{8},\nicefrac{-3}{8}),(\nicefrac{-3}{8},\nicefrac{1}{8},\nicefrac{1}{8}),(\nicefrac{1}{8},\nicefrac{-3}{8},\nicefrac{1}{8})) were used as initial guess. In the case of SLWF+C, we optimise the first WF and constrain its centre to sit at (\nicefrac{1}{4},\nicefrac{1}{4},\nicefrac{1}{4}). 
For SAWF one $s$-type and three $p$-type orbitals centred on the As atom are used for the initial guess. For all plots we choose an isosurface level of $\pm$ 0.5 \AA$^{\nicefrac{-3}{2}}$ (blue for $+$ values and red for $-$ values) using the Vesta visualisation program\cite{vesta}.
Bottom (table): Cartesian coordinates of the centres $\overbar{\bfr}$, (minimised) individual spreads $\braket{r^2}-\overbar{r}^2$ and the total spread $\Omega$ of all four valence WFs using the above mentioned four different minimisation schemes and initial guesses.
}\label{fig:slwf_sawf}
\end{figure}

Marzari and Vanderbilt\cite{MV_PRB56} have shown that the \MLWFs{} for the 4-dimensional valence manifold are centred on the four As-Ga bonds, have $sp^3$ character and can be found by specifying  four $s$-like orbitals on each covalent bond as initial guess (a representative is shown in Fig.~\ref{fig:slwf_sawf}(a)). These bond-centred functions correspond to the irreducible representation $A_{1}$ of the site-symmetry group $C_{3v}$ of the Wyckoff position {\it e}. Hence, the \MLWFs{} can also be obtained with the SAWF approach by specifying the centres and the shapes of the initial projections, e.g. four $s$-like orbitals centred on the four As--Ga bonds, and the symmetry operations in the point group $C_{3v}$. 

Using the SAWF method we can enforce the WFs to have the local site symmetries. In particular, since $T_d$ has 5 {\it irreps} of dimension 1, 1, 2, 3 and 3 respectively, one can form an 1+3--dimensional representation for the four SAWFs. 
Thus, a set of initial projections compatible with the symmetries of the valence bands is: one $s$-like orbital (1-dimensional {\it irrep} whose character is $A_1$) and three $p$-like orbitals (3-dimensional {\it irrep} whose character is $T_2$) centred on As. Fig.~\ref{fig:slwf_sawf}(b) shows the SAWF which corresponds to the $A_1$ representation and transforms like the identity under $T_d$.

The same result can be obtained with the SLWF+C method by selectively localising one function $J'=1$ $(J=4)$ and constraining its centre to sit on the As site $(\nicefrac{1}{4},\nicefrac{1}{4},\nicefrac{1}{4})$. In the case of GaAs the SLWF+C method turns out to be very robust, to the point that four $s$-like orbitals randomly centred in the unit cell can be used as initial guess without affecting the result of the optimised function. 
Fig.~\ref{fig:slwf_sawf}(c) shows the resulting function using the SLWF method without constraints, while Fig.~\ref{fig:slwf_sawf}(d) shows the result using SLWF+C.

It is worth to note that for this particular system, it is possible to achieve this result with the maximal localisation procedure if one carefully selects the initial projections, i.e., one $s$-like  and three $p$-like orbitals on the As atom. The resulting WFs will possess the local site symmetries but will not correspond to the global minimum of the spread functional $\Omega$. More precisely, they will correspond to a saddle point of $\Omega$ (unstable against small perturbations of the initial projections).
In Fig.~\ref{fig:slwf_sawf}-(a)-(b)-(c)-(d) we show a comparison of the centre and symmetries of a MLWF, SAWF, SLWF and SLWF+C; the individual spreads and total spreads---for all four valence states---are reported in the Table below it.

\subsection{Parallelisation}
In \Wannier{} v3.0 we have implemented an
efficient parallelisation scheme for the calculation of MLWFs using the message passing interface (MPI). 

\begin{figure}[t!]
\centering
{\includegraphics[width=6.5cm]{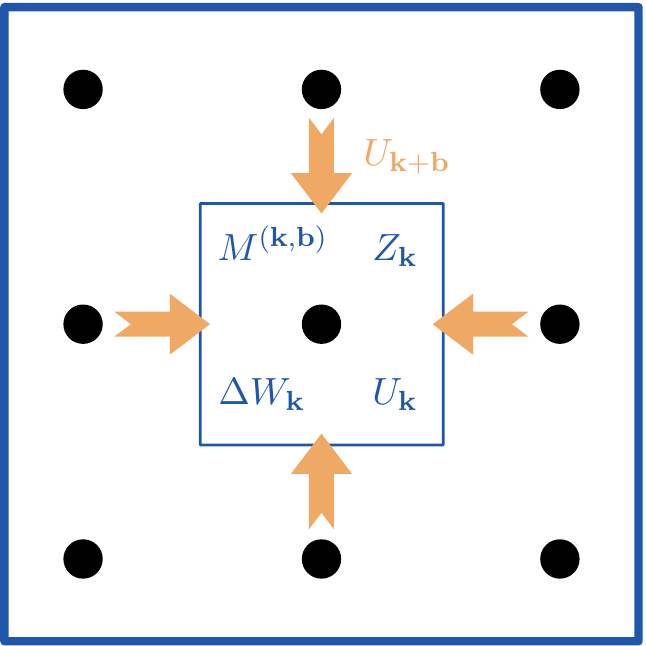}}
\caption{Illustration of the parallelisation scheme for a $3\times3$ mesh of $k$-points (black dots) and one MPI process per $k$-point. The calculation of the $\matMkb$, $Z_{\bfk}$, $\Delta W_{\bfk}$ and $\matUk$ matrices are distributed over processes by $k$-point. The $U_{\bfk+\bfb}$ matrices for the neighbouring $k$-points are sent from process to process (orange arrows) for the calculation of the $\matMkb$ and $Z_{\bfk}$ matrices.}\label{fig:mpi1}
\end{figure}

{\it Calculation of the spread.} The time-consuming part in the evaluation of the
spread $\Omega$ is updating the $\matMkb$ matrices according to Eq.~\eqref{eq:updated-Mmn}, since this requires computing overlap matrix elements between all pairs of bands, and between all $k$-points $\bfk$ and their neighbours $\bfk+\bfb$. Therefore, an efficient speed up for the evaluation of the spread can be achieved by distributing over several processes the calculation of the $\matMkb$ matrices for different $k$-points. 
In order to compute the $\matMkb$ according to Eq.~\eqref{eq:updated-Mmn}, the $U_{\bfk+\bfb}$ matrices are sent from process to process prior to the calculation of the overlap matrices. We stress the fact that the $U_{\bfk+\bfb}$ matrices are the only large arrays that have to be shared between processes, which limits the time spent in communication. 
The relatively large $\matMkb$ matrices are not sent between processes for the evaluation of Eqs.~\eqref{eq:omi-k} and \eqref{eq:omt-k}. Instead, it is enough to collect the contributions to the spread from the different $k$-points, i.e., a set of scalars, and then sum them up for evaluation of the total spread.
This parallelisation scheme is illustrated in Fig.~\ref{fig:mpi1} for a $3 \times 3$ mesh of $k$-points with 9 MPI processes.

{\it Minimisation of the spread.} The minimisation of the spread functional is based on an iterative steepest-descent or conjugate-gradient algorithm. In each iteration, the unitary matrices $\matUk$ are updated according to $\matUk = \matUk \exp{(\Delta W_{\bfk})}$~\cite{MV_PRB56}, where $\Delta W_{\bfk}=\alpha \mathcal{D}_{\bfk}$, 
see Eq.~\eqref{eq:U-update}.
Updating the $\matUk$ matrices according to this equation is by
 far the most time-consuming part in the iterative minimisation algorithm, as it requires a diagonalisation of the $\Delta W_{\bfk}$ matrices. A significant speed-up can be obtained, however, by distributing the diagonalisation of the different $\Delta W_{\bfk}$ matrices over several processes, and performing the calculations fully in parallel. The evaluation of $\Delta W_{\bfk}$ essentially requires the calculation of the overlap matrices $\matMkb$, as discussed above.
 
{\it Disentanglement.} The disentanglement procedure is concerned with finding the optimal subspace $\mathcal{S}_{\bfk}$. As the functional $\omi$ measures the global subspace dispersion across the Brillouin zone, at first sight it is not obvious that the task of minimising the spread $\omi$ can be parallelised with respect to the $k$-points. In the iterative algorithm of Eq.~\eqref{eq:disentangle1}, the systematic reduction of the spread functional at the $i^{\text{th}}$ iteration is achieved by minimising the spillage of the subspace $\mathcal{S}^{(i)}_{\bfk}$ over the neighbouring subspaces from the previous iteration $\mathcal{S}^{(i-1)}_{\bfk+\bfb}$. This problem reduces to the diagonalisation of $N$ independent matrices ($N$ is the total number of $k$-points of the mesh), where an
efficient speed-up of the disentanglement procedure can be achieved by distributing the diagonalisation of the $Z^{(i)}_{\bfk}$ matrices of Eq.~\eqref{eq:z_matrix} over several processes, which can be done fully in parallel. Since the construction of $Z^{(i)}_{\bfk}$ only requires the knowledge of the $U^{(i-1)}_{\bfk+\bfb}$ matrices, these must be communicated between processes, as shown in Fig.~\ref{fig:mpi1}. This results in a similar time spent in communication for the disentanglement part of the code as for the wannierisation part.

{\it Distribution of large matrices.} The parallelisation scheme discussed above relies on the parallel evaluation of relevant matrices over $k$-points on each processor. For systems with large number of $k$-points and bands, it is also desirable to distribute the matrices themselves among the available cores so that the memory per core required to store them is reduced. For example, in the case of isolated bands, storing all the $\matMkb$ matrices requires an allocation of dimension $J \times J \times N \times N_{b}$, where 
$N_{b}$ is the number neighbours of each of the $N$ $k$-points of the mesh. By distributing the storage across $N_{\text{c}}$ cores, the storage requirement per core decreases accordingly by a factor of approximately $N_{\text{c}}$.

\begin{figure}[t!]
\centering
{\includegraphics[width=8cm]{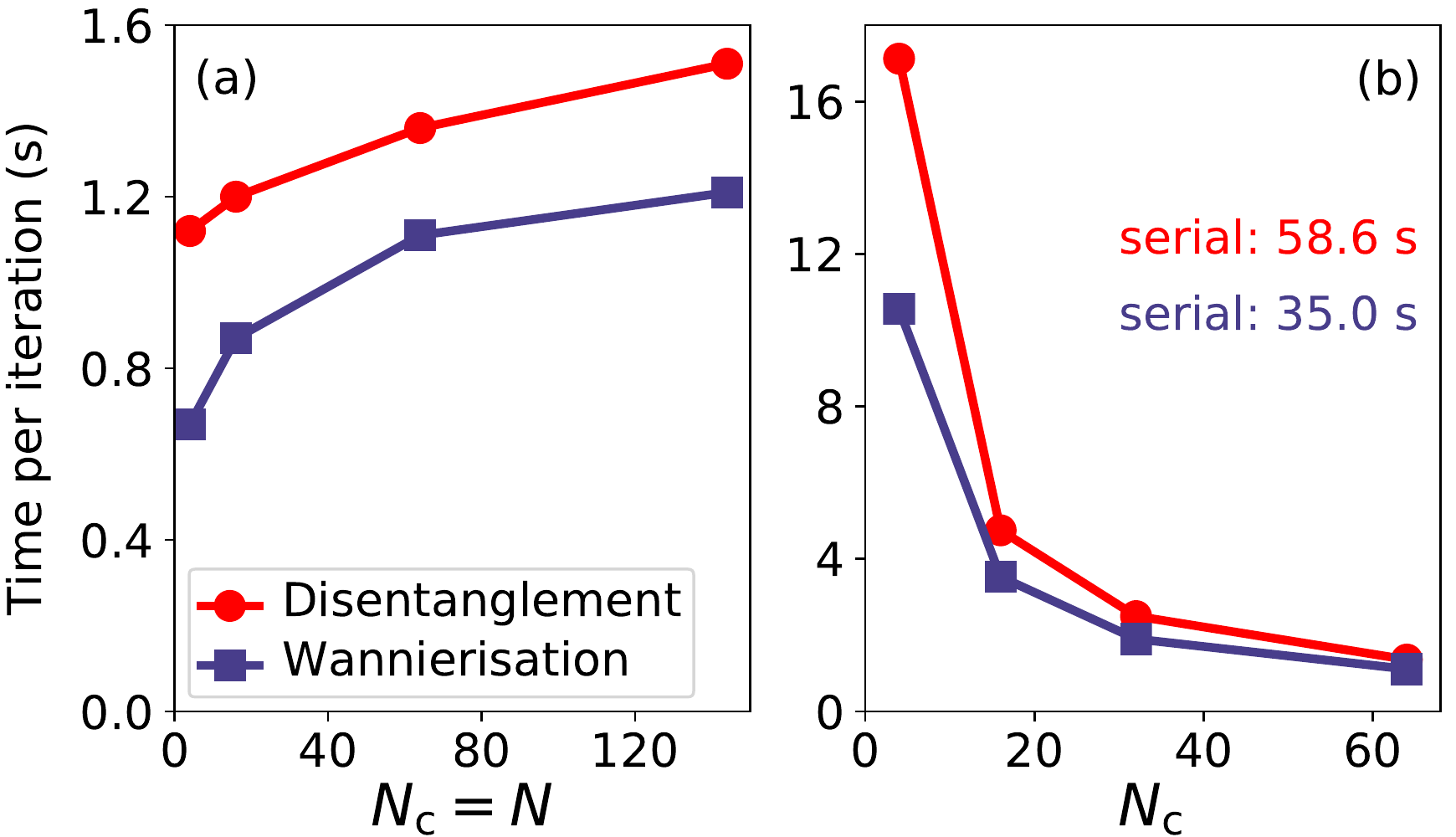}}
\caption{Plots of the time per single minimisation iteration as a function of the number of cores $N_\text{c}$. (a) Weak scaling of the implementation, where the number of $k$-points per process is fixed to one, i.e., $N_\text{c} = N$. The time only increases by a factor 1.3 (1.8) for the disentanglement (wannierisation) parts of the code, when going from $N_\text{c}=4$ to $N_\text{c}=144$. (b) Strong scaling of the algorithm for a fixed number of $k$-points $N = 64$. The time per iteration with one single CPU (serial) is reported in the figure. } 
\label{fig:mpi2}
\end{figure}

{\it Performance.} We have tested the performance of this parallelisation scheme for the calculation of the MLWFs in a $L1_0-$FePt(5)/Pt(18) thin film. Computational details were given in Ref.~\onlinecite{Geranton-PRB2015}. The benchmarks have been performed on the JURECA supercomputer of the  J\"ulich Supercomputing Center. We have extracted an optimal subspace of dimension $J = 414$ from a set of 580 Bloch states per $k$-point. 
The upper limit of the inner window was set to 5~eV above the Fermi energy, and 414 MLWFs were constructed by minimising the spread $\Omega$. The performance benchmark was based on the average wall-clock time for a single iteration of the minimisation procedure (several thousand iterations are usually needed for convergence).
We first analyse the weak scaling of our implementation, i.e., how the computation time varies with the number of cores $N_\text{c}$ for a fixed number of $k$-points per process. We show in Fig.~\ref{fig:mpi2}(a) the time per iteration for the disentanglement and wannierisation parts of the minimisation, always using one $k$-point per process. As we vary the number of $k$-points $N$ from 4 to 144, the computation time increases only by a factor of 1.3 and 1.8 for disentanglement and wannierisation, respectively.
We then demonstrate the strong scaling of our parallelisation scheme in Fig.~\ref{fig:mpi2}(b), i.e., how the computation time varies with the number of cores $N_\text{c}$ for a fixed number $N=64$ of $k$-points. When varying the number of cores from 4 to 64, we observe a decrease of the computation time per iteration by a factor of 12.6 and 9.5 for disentanglement and wannierisation, respectively. The deviation from ideal scaling is mostly explained by the time spent in inter-core communication of the $U_{\bfk+\bfb}$ matrices.

\section{Enhancements in functionality}\label{sec:enhancements}
In this section we describe a number of enhancements to the functionality of the core \Wannier{} code, namely: the ability to compute and visualise spinor-valued WFs, including developments to the interface with the \QE{} package to cover also the case of non-collinear spin calculations performed with ultrasoft pseudopotentials (previously not implemented); an improvement to the method for interpolating the $k$-space Hamiltonian; the ability to select a subset from a larger set of projections of localised trial orbitals onto the Bloch states for initialising the WFs; and new functionality for plotting WFs in Gaussian cube format on WF-centred grids with non-orthogonal translation vectors. 

\subsection{Spinor-valued Wannier functions with ultrasoft and projector-augmented-wave pseudopotentials} 
\label{sec:spinor}
The calculation of the overlap matrix in Eq.~\eqref{eq:def-Mmn} within the ultrasoft-pseudopotential formalism proceeds via the inclusion of so-called augmentation functions,\cite{Ferretti2007}
\begin{equation}
\begin{aligned}
M^{(\mathbf{k,b})}_{mn} &=  \inprod{u_{m\bfk}}{u_{n,\bfk+\bfb}} \\
&\quad + \sum_{Iij} Q^I_{ij}(\bfb)
\elm{\psi^{\mathrm{ps}}_{m\bfk}}{B^{(\bfk,\bfb)}_{Iij}}{\psi^\mathrm{ps}_{n,\bfk+\bfb}},
\label{eq:mkb-q}
\end{aligned}
\end{equation}
where $\ket{\psi^{\mathrm{ps}}_{m\bfk}}$ is the pseudo-wavefunction,
\begin{equation}
    Q^{I}_{ij}(\bfb) = \int\!\!\diffr\,\, Q^I_{ij}(\bfr) e^{-i\bfb\cdot\bfr}
\end{equation}
is the Fourier transform of the augmentation charge, and $B^{(\bfk,\bfb)}_{Iij} = \ket {\beta^{\bfk}_{Ii}}\bra
{\beta^{\bfk+\bfb}_{Ij}}$, where $\ket{\beta^{\bfk}_{Ii}}$ denotes the $i^{\rm th}$ projector of the pseudopotential on the $I^{\rm th}$ atom in the unit cell. We refer to Appendix B of Ref.~\onlinecite{Ferretti2007} for detailed expressions.

When spin-orbit coupling is included, the Bloch functions 
become two-component spinors 
$(\psi^{\uparrow}_{n \mathbf{k}}(\mathbf{r}), \psi^{\downarrow}_{n \mathbf{k}}(\mathbf{r}))^\textrm{T}$, where $\psi^{\sigma}_{n \mathbf{k}}(\mathbf{r})$ is the spin-up (for $\sigma=\uparrow$) or spin-down (for $\sigma=\downarrow$) component with respect to the chosen spin quantisation axis. 
Accordingly, $Q^I_{ij}(\bfb)$ becomes $Q_{ij}^{I\sigma\sigma'}(\bfb)$ (see Eq.~(18) in Ref.~\onlinecite{DalCorso2005}) and 
\Eqref{eq:mkb-q} becomes
\begin{equation}
\begin{aligned}
M^{(\mathbf{k,b})}_{mn} & = \braket{u_{m\bfk}\vert u_{n,\bfk+\bfb}} \\ 
 &\quad + \sum_{Iij\sigma\sigma'} Q^{I\sigma\sigma'}_{ij}(\bfb)
\elm{\psi^{\mathrm{ps},\sigma}_{m\bfk}}{B^{(\bfk,\bfb)}_{Iij}}  {\psi^\mathrm{ps,\sigma'}_{n,\bfk+\bfb}}.
\end{aligned}
\end{equation}
The above expressions, together with the corresponding ones for  the matrix elements of the spin operator, have been implemented in the \pwtowan{} interface between \QE{}
and \Wannier{}.

The plotting routines of \Wannier{} have also been adapted to work with the complex-valued spinor WFs obtained from calculations with spin-orbit coupling.
It then becomes necessary to decide how to represent graphically the information contained in the two spinor components. 

One option is to only plot the norm $|\psi _{n \mathbf{k}}(\mathbf{r})|=\sqrt{|\psi^{\uparrow}_{n \mathbf{k}}(\mathbf{r})|^2+|\psi^{\downarrow}_{n \mathbf{k}}(\mathbf{r})|^2}$ of spinor WFs, which is reminiscent of the total charge density in the case of a 2$\times$2 density matrix in non-collinear DFT.
Another possibility is to plot independently the up- and down-spin components of the spinor WF. Since each of them is in general complex-valued, two options are provided in the code: 
(i) to plot only the magnitudes $|\psi^{\uparrow}_{n \mathbf{k}}(\mathbf{r})|$ and $|\psi^{\downarrow}_{n \mathbf{k}}(\mathbf{r})|$ of the two components; or (ii) to encode the phase information by outputting $|\psi^{\uparrow}_{n \mathbf{k}}(\mathbf{r})|\text{sgn}(\text{Re}\{\psi^{\uparrow}_{n \mathbf{k}}(\mathbf{r})\})$ and $|\psi^{\downarrow}_{n \mathbf{k}}(\mathbf{r})|\text{sgn}(\text{Re}\{\psi^{\downarrow}_{n \mathbf{k}}(\mathbf{r})\})$, where $\text{sgn}$ is the sign function. Which of these various options is adopted by the \Wannier{} code is controlled by two input parameters, \inputvar{wannier\_plot\_spinor\_mode} and \inputvar{wannier\_plot\_spinor\_phase}.

Finally we note that, for WFs constructed from ultrasoft pseudopotentials or within the projector-augmented-wave (PAW) method, only pseudo-wavefunctions represented on the soft FFT grid are considered in plotting WFs within the present scheme, that is, the WFs are not normalised.

\subsection{Improved Wannier interpolation by minimal-distance replica selection}
The interpolation of band structures (and many other quantities) based on Wannier functions is an extremely powerful tool\cite{Lee_PRL05,wang-prb06,Yates_PRB07}. In many respects it resembles Fourier interpolation, which uses discrete Fourier transforms to reconstruct faithfully continuous signals 
from a discrete sampling, provided that the signal has a finite bandwidth and that the sampling rate is at least twice the bandwidth (the so-called Nyquist--Shannon condition).

In the context of Wannier interpolation, the ``sampled signal'' is the set of matrix elements 
\begin{equation}\label{eq:H-ki}
    H_{mn\bfk_j}=\elm{\chi_{m\bfk_j}}{H}{\chi_{n\bfk_j}}
\end{equation}
of a lattice-periodic operator such as the Hamiltonian,
defined on the same uniform grid $\{\bfk_j\}$ that was used to minimise the Wannier spread functional (see Sec.~\ref{sec:isolated}).
The states $\ket{\chi_{n\bfk_j}}$ are the Bloch sums of the WFs,
related to \textit{ab initio} Bloch eigenstates by $\ket{\chi_{n\bfk_j}}=\sum_m\ket{\psi_{m\bfk_j}}U_{mn\bfk_j}$.

To reconstruct the ``continuous signal'' $H_{nm\bfk}$ at arbitrary $\bfk$, the matrix elements of \Eqref{eq:H-ki} are first mapped onto real space using the discrete Fourier transform
\begin{equation}\label{eq:twosites}
\widetilde{H}_{mn\bfR} = \braket{w_{m\mathbf{0}}|H|w_{n\bfR}} =   \frac{1}{N}\sum_{j=1}^{N} e^{-i\bfk_j\cdot\bfR} H_{mn\bfk_j},
\end{equation}
where $N=N_1\times N_2 \times N_3$ is the grid size (which is also the number of $k$-points in \Wannier{}).
The matrices $H_{mn\bfk_j}$ are then interpolated onto an arbitrary $\bfk$ using an inverse discrete Fourier transform,
\begin{equation}\label{eq:simple_interpolation}
H_{mn\bfk} =   \sum_{\bfR'} e^{i\bfk\cdot\bfR'} \widetilde{H}_{mn\bfR'}\,,
\end{equation}
where the sum is over $N$ lattice vectors $\bfR'$, and the interpolated energy eigenvalues are obtained by diagonalising $H_{\bfk}$. In the limit of an infinitely dense grid of $k$-points the procedure is exact and the sum in Eq.~\eqref{eq:simple_interpolation} becomes an infinite series. Owing to the real-space localisation of the Wannier functions, the matrix elements $\widetilde{H}_{mn\bfR}$ become vanishingly small when the distance between the Wannier centres exceeds a critical value $L$ (the ``bandwidth'' of the Wannier Hamiltonian), so that actually only a finite number of terms contributes significantly to the sum in Eq.~\eqref{eq:simple_interpolation}.  This means that, even with a finite $N_1\times N_2 \times N_3$ grid, the interpolation is still accurate provided that --~by analogy with the Nyquist--Shannon condition~-- the ``sampling rate'' $N_i$ along each cell vector $\mathbf{a}_i$ is sufficiently large to ensure that $N_i |\mathbf{a}_i|>2L$.

Still, the result of the interpolation crucially depends on the choice of the $N$ lattice vectors to be summed over in \Eqref{eq:simple_interpolation}. Indeed, when using a finite grid, there is a considerable freedom in choosing the set $\{\bfR'\}$ as  $\widetilde{H}_{mn\bfR}$ is invariant under $\bfR\rightarrow \bfR+\mathbf{T}$ for any vector $\mathbf{T}$ of the
Born--von Karman superlattice
generated by $\{\mathbf{A}_i = N_i\mathbf{a}_i\}$.
The phase factor in \Eqref{eq:simple_interpolation} is also invariant when $\bfk\in\{ \bfk_j\}$, but not
for arbitrary $\bfk$. Hence we need to choose, among the infinite set of ``replicas'' $\bfR'=\bfR+\mathbf{T}$ of $\bfR$, which one to include in Eq.~\eqref{eq:simple_interpolation}.
We take the original vectors $\bfR$ to lie within the Wigner--Seitz supercell centred at the origin. If some of them fall on its boundary then their total number exceeds $N$ and weight factors must be introduced in \Eqref{eq:simple_interpolation}. For each combination of $m$, $n$ and $\bfR$,
the optimal choice of $\mathbf{T}$ is the one that minimises the distance
\begin{equation}\label{eq:d-R-T}
    |\bfr_m - (\bfr_n+\bfR+\mathbf{T})|
\end{equation}
between the two Wannier centres. With this choice, the spurious effects arising from the artificial supercell periodicity are minimised.

\begin{figure}
\includegraphics[width=8cm]{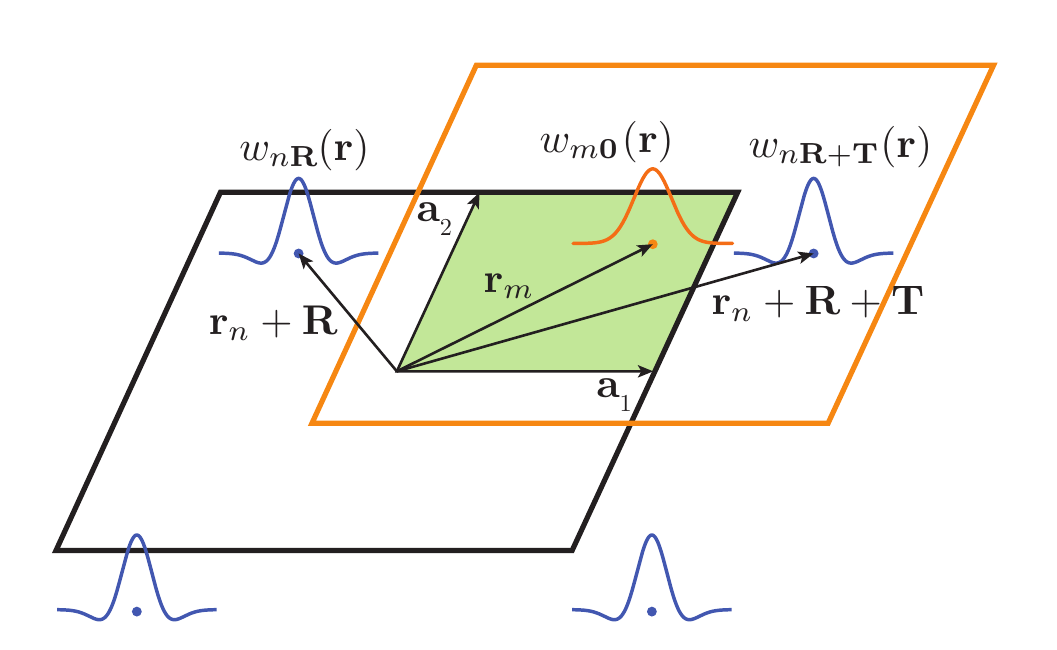}
\caption{Owing to the periodicity of the Wannier functions over the Born-von Karman supercell (with size $2\times2$ here), the matrix element $\widetilde{H}_{mn\bfR}$ describes the interaction between the $m^{\rm th}$ WF $w_{m\mathbf{0}}$ (shown in orange) with centre $\bfr_m$
inside the home unit cell $\bfR=\mathbf{0}$ (green shaded area) and the $n^{\rm th}$ WF $w_{n\bfR}$ 
(shown in blue) centred inside the unit cell $\bfR$, or any of its supercell-periodic replicas displaced by a superlattice vector $\mathbf{T}$. When performing Wannier interpolation, we now impose a minimal-distance condition by choosing the replica $w_{n,\bfR+\mathbf{T}}$ of $w_{n\bfR}$  whose centre lies within the Wigner--Seitz supercell centred at $\bfr_m$ (thick orange line).
\label{fig:ws_distance}}
\end{figure}

Earlier versions of \Wannier{} implemented a simplified procedure whereby the vectors $\bfR'$ in 
\Eqref{eq:simple_interpolation} were chosen to coincide with the unshifted vectors $\bfR$ that are closer to the origin than to any other point $\mathbf{T}$ on the superlattice, irrespective of the WF pair $(m,n)$. 
As illustrated in Fig.~\ref{fig:ws_distance}, this procedure does not always lead to the shortest distance between the pair of WFs, especially when some of the $N_i$ are small and the Wannier centres are far from the origin of the cell.

 \Wannier{} now implements an improved algorithm that enforces the minimal-distance condition of \Eqref{eq:d-R-T}, yielding a more accurate Fourier interpolation. The algorithm is the following:
\begin{enumerate}
\item[(a)] For each term in \Eqref{eq:simple_interpolation} pick, among all the replicas $\bfR'=\bfR+\mathbf{T}$ of $\bfR$,  the one that minimises 
the distance between Wannier centres (\Eqref{eq:d-R-T}).
\item[(b)] If there are $\mathcal{N}_{mn\bfR}$ different vectors  $\mathbf{T}$ for which the distance of \Eqref{eq:d-R-T} is minimal, then include all of them in \Eqref{eq:simple_interpolation} with a weight factor $1/\mathcal{N}_{mn\bfR}$.
\end{enumerate}

An equivalent way to describe these steps is that (a) we choose $\mathbf{T}$ such that $\bfr_n+\bfR+\mathbf{T}$ falls inside the Wigner--Seitz supercell centred at $\bfr_m$ (see Fig.~\ref{fig:ws_distance}), and that (b) if it falls on a face, edge or vertex of the Wigner--Seitz supercell, we  keep all the equivalent replicas with an appropriate weight factor. In practice the condition in step~(b) is enforced within a certain tolerance, to account for the numerical imprecision 
in the values of the Wannier centres and in the definition of the unit cell vectors. Although step~(b) is much less important than (a) for obtaining a good Fourier interpolation, it helps ensuring that the interpolated bands respect the symmetries of the system; if step~(b) is skipped, small artificial band splittings may occur at high-symmetry points, lines, or planes in the BZ.

\begin{figure}
\includegraphics[width=8cm]{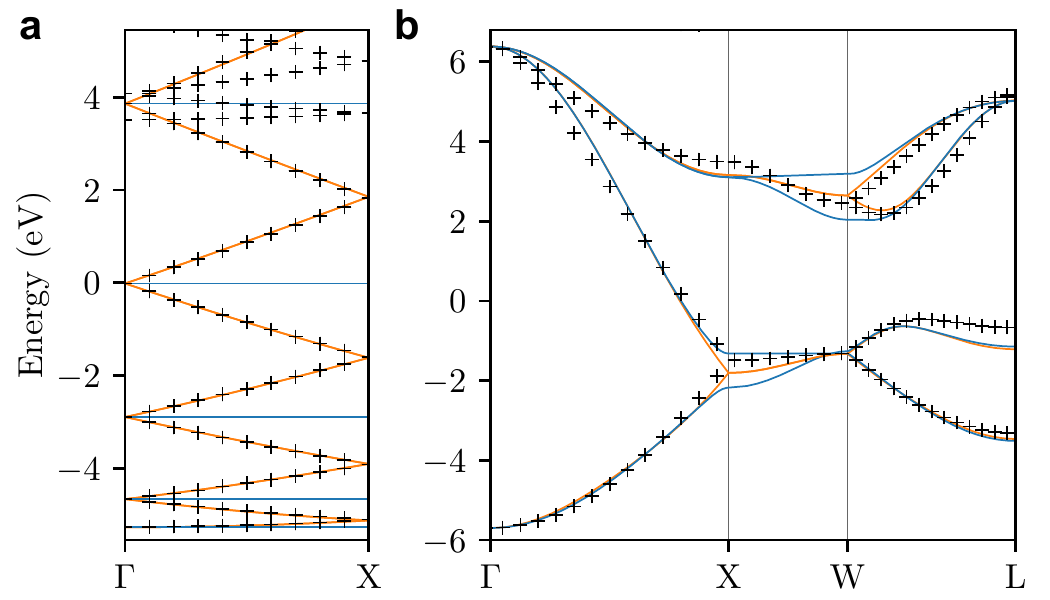}\caption{Comparison between the  bands obtained using the earlier interpolation procedure (blue lines), those obtained using the (current) modified approach of \Eqref{eq:backwithweight} (orange lines), and the \textit{ab initio} bands (black crosses). (a) Linear chain of carbon atoms, with 12 atoms per unit cell (separated by a distance of 1.3 \AA\ along the $z$ direction) and $\Gamma$-point sampling. 36 Wannier functions have been computed starting from projections over $p_x$ and $p_y$ orbitals on carbon atoms and  $s$-orbitals midbond between them. A frozen window up to the Fermi energy (set to zero in the plot) has been considered, while the disentanglement window included all states up to $\sim14$~eV above the Fermi level.
 (b) Bulk silicon, with the BZ sampled on an unconverged $3\times 3 \times 3$ grid of $k$-points.}\label{fig:ws_bands}
\end{figure}

The procedure outlined above amounts to replacing \Eqref{eq:simple_interpolation} with
\begin{equation}
    H_{mn\bfk} = \sum_{\bfR}
    \frac{1}{\mathcal{N}_{mn\bfR}}
    \sum_{j=1}^{\mathcal{N}_{mn\bfR}} e^{i \bfk \cdot (\bfR+\mathbf{T}^{(j)}_{mn\bfR})}
      \widetilde{H}_{mn\bfR} \,,\label{eq:backwithweight}
\end{equation}
where $\{\mathbf{T}^{(j)}_{mn\bfR}\}$ are the $\mathcal{N}_{mn\bfR}$ vectors $\mathbf{T}$ that minimise the distance of \Eqref{eq:d-R-T} for a given combination of $m$, $n$ and~$\bfR$; $\bfR$ lies within the Wigner--Seitz supercell centred on the origin.

The benefits of this modified interpolation scheme  are most evident when considering a large unit cell sampled at the $\Gamma$ point only. In this case $N=1$ so that \Eqref{eq:simple_interpolation} with $\{\bfR'\}=\{\bfR\}=\{\mathbf{0}\}$ would reduce to $H_{mn\bfk} =  \widetilde{H}_{mn\mathbf{0}}$, yielding interpolated  bands that do not disperse with $\bfk$. This is nonetheless an artefact of the choice $\{\bfR'\}=\{\mathbf{0}\}$ (of earlier versions of \Wannier{}) and not an intrinsic limitation of Wannier interpolation, as first demonstrated in Ref.~\onlinecite{Lee_PRL05} for one-dimensional systems. Indeed, equation~\eqref{eq:backwithweight}, which in a sense extends Ref.~\onlinecite{Lee_PRL05} to any spatial dimension, becomes in this case
\begin{equation}
    H_{mn\bfk} = \frac{\widetilde{H}_{mn\mathbf{0}}}{\mathcal{N}_{mn\mathbf{0}}}
    \sum_{j=1}^{\mathcal{N}_{mn\mathbf{0}}} e^{i \bfk \cdot \mathbf{T}^{(j)}_{mn\mathbf{0}}}\,,
\end{equation}
which can produce dispersive bands. 
This is illustrated in Fig.~\ref{fig:ws_bands}(a) for the case of a one-dimensional chain of carbon atoms: the interpolated bands obtained from \Eqref{eq:simple_interpolation}  with $\{\bfR'\}=\{\bfR\}=\{\mathbf{0}\}$ (earlier version of \Wannier{}) are flat, while those obtained from \Eqref{eq:backwithweight} (new versions of \Wannier{}) are in much better agreement with the dispersive \emph{ab initio} bands up to a few eV above the Fermi energy.

Clear improvements in the interpolated bands are also obtained for
bulk solids, as shown in Fig.~\ref{fig:ws_bands}(b) for the case of silicon. The earlier implementation breaks the two-fold degeneracy along the X$-$W line, with one of the two bands becoming flat. The new procedure recovers the correct degeneracies, and reproduces more closely the \emph{ab initio} band structure (the remaining small deviations are due to the use of a coarse $k$-point mesh that does not satisfy the Nyquist--Shannon condition, and would disappear for denser $k$-grids together with the differences between the two interpolation procedures).

\subsection{Selection of projections} 
In many cases, and particularly for entangled bands, it is necessary to have a good initial guess for the MLWFs in order to properly converge the spread to the global minimum.
Determining a good initial guess often involves a trial and error approach, using different combinations of orbital types, orientations and positions.
While for small systems performing many computations of the projection matrices is relatively cheap, for large systems there is a cost associated with storing and reading the wavefunctions to compute new projection matrices for each new attempt at a better initial guess.
Previously, the number of projections that could be specified had to be equal to the number $J$ of WFs to be constructed. The latest version of the code lifts this restriction,
  making it possible to define in the pre-processing step a larger number $J_{+}>J$ of projection functions to consider as initial guesses.  In this way, the computationally expensive and potentially I/O-heavy construction of the projection matrices $\matAk$ is performed only once for all possible projections that a user would like to consider. 

Once the $\matAk$ matrices (of dimension $J\times J_{+}$ at each $\bfk$) have been obtained, one proceeds with constructing the MLWFs by simply selecting, via a new input parameter (\inputvar{select\_projections}) of the \Wannier{} code, which $J$ columns to use among the $J_{+}$ that were computed by the interface code.
Experimenting with different trial orbitals can thus be achieved by simply selecting a different set of  projections within the \Wannier{} input file, without the need to perform the pre-processing step again.

Similarly, another use case for this new option is the construction of WFs for the same material but for different groups of bands.
Typically one would have to modify the \Wannier{} input file and run the interface code multiple times, while now the interface code may compute $\matAk$ for a superset of trial orbitals just once, and then different subsets may be chosen by simple modification of a single input parameter. As a demonstration, we have adapted \texttt{example11} of the \Wannier{} distribution (silicon band structure), that considers two band groups: (a)
  the valence bands only, described by four bond-centred $s$ orbitals, and (b) the four valence and the four lowest-lying conduction bands together, described by atom-centred $sp^3$ orbitals.
In the example, projections onto all 12 trial orbitals are provided, and the different cases are covered by specifying in the \Wannier{} input file which subset of projections is required.

\subsection{Plotting cube files with non-orthogonal vectors} 
In \Wannier{} v3.0 it is possible to plot the \MLWFs{} in real-space in Gaussian cube format, including the case of non-orthogonal cell lattice vectors. Many modern visualisation programs such as Vesta\cite{vesta} are capable of handling non-orthogonal cube files and the cube file format can be read by many computational chemistry programs. \Wannier{}'s representation of \MLWFs{} in cube format can be significantly more compact than using the alternative {\tt xsf} format. With the latter, \MLWFs{} are calculated (albeit with a coarse sampling) on a supercell of the computational cell that can be potentially large (the extent of the supercell is controlled by an input parameter \inputvar{wannier\_plot\_supercell}). Whereas, with the cube format, each Wannier function is represented on a grid that is centred on the Wannier function itself and has a user-defined extent, which is the smallest parallelepiped (whose sides are aligned with the cell vectors) that can enclose a sphere with a user-defined radius \inputvar{wannier\_plot\_radius}. Because \MLWFs{} are strongly localised in real space, relatively small cut-offs are all that is required, significantly smaller than the length-scale over which the \MLWFs{} themselves are periodic. As a result, the cube format is particularly useful when a more memory-efficient representation is needed. The cube format can be activated by setting the input parameter \inputvar{wannier\_plot\_mode} to \inputvar{cube}, and the code can handle both isolated molecular systems (treated within the supercell approximation) as well as periodic crystals by setting \inputvar{wannier\_plot\_mode} to either \inputvar{molecule} or \inputvar{crystal}, respectively.

\section{New post-processing features}\label{sec:postproc}
Once the electronic bands of interest have been disentangled and wannierised to obtain well-localised WFs, the \Wannier{} software package includes a number of modules and utilities that use these WFs to calculate various electronic-structure properties. Much of this functionality exists within \exec{postw90}, an MPI-parallel code that forms an integral part of the \Wannier{} package. In v2.x of \Wannier{}, \exec{postw90} included functionality for computing densities of states and partial densities of states, energy bands and Berry curvature along specified lines and planes in $k$-space, anomalous Hall conductivity, orbital magnetisation and optical conductivity, Boltzmann transport coefficients within the relaxation time approximation, and band energies and derivatives on a generic user-defined list of $k$-points. Some further functionality exists in a set of utilities that are provided as part of the \Wannier{} package, including a code (\module{w90pov}) to plot WFs rendered using the Persistence of Vision Raytracer (POV-Ray)\cite{povray} code and to compute van der Waals interactions with WFs (\module{w90vdw}). 

In addition, there are a number of external packages for computing advanced properties based on WFs and which interface to \Wannier{}. These include codes to generate tight-binding models such as \package{pythTB}~\cite{pythtb} and \package{tbmodels}~\cite{tbmodels}, quantum transport codes such as \package{sisl}~\cite{zerothi_sisl}, \package{gollum}~\cite{gollum}, \package{omen}~\cite{omen} and \package{nanoTCAD-ViDES}~\cite{nanotcadvides}, the \package{EPW}~\cite{epw} code for calculating properties related to electron-phonon interactions and \package{WannierTools}~\cite{WannierTools} for the investigation of novel topological materials.

Below we describe some of the new post-processing features of \Wannier{} that have been introduced in the latest version of the code, v3.0.

\subsection{postw90.x: Shift Current}
\label{sec:shift-current}
The photogalvanic effect (PGE) is a nonlinear optical response
that consists in the generation of a direct current (DC) when light is
absorbed.\cite{belinicher-spu80,sturman-book92,ivchenko-book97} It can
be divided phenomenologically into linear (LPGE) and circular (CPGE)
effects, which have different symmetry requirements within the
acentric crystal classes. The CPGE requires elliptically-polarised
light, and occurs in gyrotropic crystals (see next subsection).  The
LPGE occurs with linearly or unpolarised light as well; it is present in
piezoelectric crystals and is given by
\begin{equation}
\label{eq:LPVE}
J_a(0)=2\sigma_{abc}(0;\omega,-\omega)E_b(\omega)E_c(-\omega),
\end{equation}
where ${\bf J}(0)$ is the induced DC photocurrent density, ${\bf E}(\omega)=\bf E^*(-\omega)$
is the amplitude of the optical electric field, and $\sigma_{abc}=\sigma_{acb}=\sigma_{abc}^*$ is a nonlinear photoconductivity tensor.

The shift current is the part of the LPGE photocurrent generated by interband light absorption.\cite{tan-cm16} Intuitively, it arises
from a coordinate shift accompanying the photoexcitation of electrons
from one band to another. Like the intrinsic anomalous Hall
effect~\cite{nagaosa-rmp10}, the shift current involves off-diagonal
velocity matrix elements between occupied and empty bands, depending not only on their magnitudes but also on their
phases~\cite{baltz-prb81,belinicher-jetp82,kristoffel-zpb82,sipe-prb00}.

The shift current along direction $a$ induced by light that is linearly polarised along $b$ is described by the following photoconductivity tensor:\cite{sipe-prb00,fregoso-prb17}
\begin{eqnarray}
\label{eq:shift-abb}
\sigma_{abb}^{\rm shift}(0;\omega,-\omega)&=&-\frac{\pi |e|^3}{\hbar^2}
\intBZ{\frac{\diffk}{(2\pi)^3}}\sum_{n,m}f_{nm\bfk}\,
R^{ab}_{nm\bfk}\nonumber\\
& &\times\,\left| r^b_{nm\bfk}\right|^2\delta(\omega_{mn\bfk}-\omega).
\end{eqnarray}
Here, $f_{nm\bfk}=f_{n\bfk}-f_{m\bfk}$ is the difference between
occupation factors, $\hbar\omega_{mn\bfk}=\epsilon_{m\bfk}-\epsilon_{n\bfk}$ is the difference between energy eigenvalues of the Bloch bands,
$r^b_{nm\bfk}$ is the $b^{\rm th}$ Cartesian component of the interband dipole matrix (the off-diagonal part of the Berry connection matrix
${\bf A}_{nm\bfk}=i\langle u_{n\bfk}\vert\partial_{\bfk} u_{m\bfk}\rangle$), and
\begin{equation}
\label{eq:shift-vector}
R^{ab}_{nm\bfk}=\partial_{k_a}\arg\left(r^b_{nm\bfk}\right)-A^a_{nn\bfk}
+A^a_{mm\bfk}
\end{equation}
is the {\it shift vector} (not to be confused with the lattice vector ${\bf R}$, or with the matrix $R^{({\bf k},{\bf b})}$ defined in Eq.~\eqref{eq:Rmn}). The shift vector has units of length, and it describes the
real-space shift of wavepackets under photoexcitation.

The numerical evaluation of Eq.~\eqref{eq:shift-vector} is tricky
because the individual terms therein are gauge-dependent, and only
their sum is unique.  Different strategies were discussed in the early
literature in the context of model
calculations\cite{kristoffel-zpb82,presting-pssb82} and more recently
for {\it ab initio} calculations.  The {\it ab initio} implementation
of Young and Rappe\cite{young-prl12} employed a gauge-invariant
$k$-space discretisation of Eq.~\eqref{eq:shift-vector}, inspired by
the discretised Berry-phase formula for electric
polarisation.\cite{king-smith-prb93} 

The implementation in \Wannier{} is based instead on the formulation of Sipe and
co-workers.\cite{aversa-prb95,sipe-prb00}  In this formulation, the shift (interband) contribution to the LPGE tensor in Eq.~\eqref{eq:LPVE} is expressed as 
\begin{equation}
\label{eq:shift-abc}
\begin{split}
\sigma^{\rm shift}_{abc}(0;\omega,-\omega)=&\frac{i\pi |e|^3}{4\hbar^2}
\intBZ{\frac{\diffk}{(2\pi)^3}}
\sum_{n,m}f_{nm\bfk}\\
&\times
\left(r^b_{mn\bfk}r^{c;a}_{nm\bfk} + r^c_{mn\bfk}r^{b;a}_{nm\bfk}\right)\\
&\times 
\left[
  \delta(\omega_{mn\bfk}-\omega)+\delta(\omega_{nm\bfk}-\omega)
\right],
\end{split}
\end{equation}
where  
\begin{equation}
\label{eq:gen-der}
r^{b;a}_{nm\bfk}=\partial_{k_a} r^b_{nm\bfk}
-i\left(A^a_{nn\bfk}-A^a_{mm\bfk}\right)r^b_{nm\bfk}
\end{equation}
is the {\it generalised derivative} of the interband dipole. When $b=c$, Eq.~\eqref{eq:shift-abc} becomes equivalent to Eq.~\eqref{eq:shift-abb}.\cite{sipe-prb00}

The generalised derivative $r^{b;a}_{nm\bfk}$ is a
well-behaved (covariant) quantity under gauge transformation but --~as in
the case of the shift vector~-- this is not the case for the individual
terms in Eq.~\eqref{eq:gen-der}, leading to numerical
instabilities. To circumvent this problem, Sipe and co-workers used
$\bfk\cdot\mathbf{p}$ perturbation theory to recast Eq.~\eqref{eq:gen-der} as a
summation over intermediate virtual states where the individual terms
are gauge-covariant.\cite{aversa-prb95,sipe-prb00} That strategy has
been successfully employed to evaluate the shift-current spectrum
from first principles.\cite{nastos-prb06,rangel-prl17}

As it is well known, similar ``sum rule'' expressions can be written for
other quantities involving $\bfk$ derivatives, such as the inverse
effective-mass tensor and the Berry curvature tensor. When evaluating such
expressions, a sufficiently large number of virtual states must be
included to achieve convergence. Alternatively, one can work with a
basis spanning a finite number of bands, such as a tight-binding or
Wannier basis, and carefully reformulate $\bfk\cdot\mathbf{p}$ perturbation theory within that
incomplete basis to avoid truncation errors. This was done first for the inverse effective-mass
tensor\cite{graf-prb95,boykin-prb95} and later for the Berry
curvature,\cite{wang-prb06} and is at the heart of the
Wannier-interpolation technique for calculating the intrinsic anomalous Hall
conductivity (AHC).\cite{wang-prb06}

A truncation-free tight-binding sum rule for the generalised
derivative of \Eqref{eq:gen-der}  was given in
Ref.~\onlinecite{cook-nc17}. Contrary to the inverse effective-mass tensor,
which only depends on the Hamiltonian matrix
elements,\cite{graf-prb95,boykin-prb95} the generalised derivative
also depends --~in some cases rather strongly~-- on the intracell
coordinates of the basis orbitals.~\cite{cook-nc17} Building on that
formulation, Wannier interpolation schemes for calculating the shift
current were recently introduced\cite{wang-prb17,azpiroz-prb18} (the
implementation in \Wannier{} follows
Ref.~\onlinecite{azpiroz-prb18}). In addition to the Hamiltonian
matrix elements and Wannier centres, the shift current was found to
depend sensitively on the off-diagonal position matrix elements.
 
The generalised derivative can be used to evaluate
other nonlinear optical responses, such as second-harmonic
generation.\cite{sipe-prb00,wang-prb17} While these are not currently
implemented in \Wannier{}, it should be straightforward to adapt the
shift-current routines for that purpose.

\subsection{postw90.x: Gyrotropic module}
The spontaneous magnetisation of ferromagnets endows their
linear conductivity tensor $\sigma_{ab}(\omega)$ with an antisymmetric part.
At $\omega=0$ that part describes the anomalous Hall conductivity
(AHC), and at finite frequencies it gives rise to magneto-optical
effects such as Faraday rotation in transmission and magnetic
circular dichroism in absorption. In paramagnets, those effects appear
under an external magnetic field.

Interestingly, an antisymmetric conductivity can be induced in certain
nonmagnetic (semi)conductors by purely electrical means, namely, by passing
a current through the sample~\cite{baranova-oc77,ivchenko-jetp78}.
Symmetry arguments indicate that this is allowed in the {\it
  gyrotropic} crystal classes, a subset of the acentric crystal
classes that includes those that are chiral, polar, or optically
active.\cite{belinicher-spu80} The first experimental demonstration
consisted in the measurement of a current-induced change in the
rotatory power of $p$-doped trigonal
tellurium.\cite{vorobev-jetp79,shalygin-pss12} The DC or
transport limit of this {\it current-induced Faraday effect} is the
current-induced anomalous Hall effect (AHE), which has become known in the recent literature
as the {\it nonlinear
  AHE}~\cite{sodemann-prl15,zhang-prb18,zhang-2dmater18,you-prb18,ma-nat19}.
Like the linear (spontaneous) AHE in ferromagnetic metals, the nonlinear
(current-induced) AHE in gyrotropic conductors has an intrinsic
contribution associated with the Berry curvature in momentum
space.\cite{sodemann-prl15}

Along with nonlinear magneto-optical and anomalous Hall effects, the flow of
electrical current in a gyrotropic conducting medium also generates a
net magnetisation. This {\it kinetic magnetoelectric effect} was
originally proposed for bulk chiral
conductors,~\cite{ivchenko-jetp78,levitov-jetp85} and later for
two-dimensional (2D) inversion layers with an out-of-plane polar
axis~\cite{edelstein-ssc90,aronov-jetp89}, where it has been studied
intensively~\cite{ganichev-book12}.  The kinetic magnetoelectric
effect in 2D --~also known as the {\it Edelstein effect}~-- is a
purely spin effect, whereas in bulk crystals an orbital contribution
is also present.~\cite{levitov-jetp85} The orbital kinetic
magnetoelectric effect was recently formulated in terms of the
intrinsic orbital moment of the Bloch
electrons,\cite{yoda-sr15,zhong-prl16} a quantity closely related to
the Berry curvature.

Another phenomenon characteristic of gyrotropic crystals is the {\it
  circular photogalvanic effect} (CPGE) that was mentioned briefly in Sec.~\ref{sec:shift-current}. This nonlinear optical effect
consists in the generation of a photocurrent that reverses sign with
the helicity of
light~\cite{ivchenko-jetp78,asnin-jetp78,belinicher-spu80,sturman-book92,ivchenko-book97},
and it occurs when light is absorbed via interband or intraband
scattering processes.  The intraband contribution to the CPGE is
closely related to the nonlinear AHE, as both arise from the Berry
curvature of the conduction
electrons~\cite{deyo-arxiv09,moore-prl10,sodemann-prl15}.

The gyrotropic effects listed above are being very actively investigated in
connection with novel materials ranging from topological
semimetals~\cite{juan-natcomms17,zhang-prb18,flicker-prb18} 
to monolayer and bilayer transition-metal
dichalcogenides~\cite{zhang-2dmater18,you-prb18,ma-nat19}.
The sensitivity of both the Berry curvature and the intrinsic orbital moment to
the details of the electronic structure, together with the need to
sample them on a dense mesh of $k$-points, calls for the development of
accurate and efficient {\it ab initio} methodologies for this class of
problems.

Building on existing Wannier-interpolation schemes for calculating the
spontaneous intrinsic AHC and orbital magnetisation,\cite{wang-prb06,lopez-prb12} 
the corresponding
methodology for gyrotropic effects was presented in
Ref.~\onlinecite{tsirkin-prb18}, where it was applied to $p$-doped
trigonal tellurium (in that work, only the intraband contribution to
the CPGE was considered). The resulting computer code has been
incorporated in \Wannier{} as the \module{gyrotropic} module.

The central task of that module is to evaluate response tensors such
as the ``Berry-curvature dipole''~\cite{sodemann-prl15}
\begin{equation}
\label{eq:gyro:D_ab}
D_{ab}=\intBZ{\frac{\diffk}{(2\pi)^3}} \sum_n
\frac{\partial \epsilon_{n\bfk}}{\partial{k_a}}
\Omega_{n\bfk}^b
\left(-\frac{\partial f_0}{\partial \epsilon}\right)_{\epsilon=\epsilon_{n\bfk}},
\end{equation}
where $f_0$ is the equilibrium occupation factor
and ${\bm\Omega}_{n\bfk}=\boldsymbol{\nabla}_{\bfk}\times{\bf A}_{nn\bfk}$ is the Berry
curvature of the Bloch bands
(the curl of the band-diagonal Berry connection ${\bf A}_{nn\bfk}$ introduced in Sec.~\ref{sec:shift-current}). 
Also of interest is the tensor $K_{ab}$, obtained by
replacing the Berry curvature in Eq.~\eqref{eq:gyro:D_ab} with the
intrinsic magnetic moment
of the Bloch states.\cite{yoda-sr15,zhong-prl16,tsirkin-prb18} 

The two tensors $D_{ab}$ and $K_{ab}$ describe several of the
aforementioned gyrotropic effects as follows:
\begin{itemize}

\item {\it intraband CPGE:}
$j_a\propto
\frac{\omega\tau^2D_{ab}}{1+\omega^2\tau^2} {\rm Im}\left[{\bf
    E}(\omega)\times{\bf E}^*(\omega)\right]_b$,

\item{\it nonlinear AHE:}
$j_a
\propto \tau
\varepsilon_{adc}D_{bd}E_bE_c$,

\item{\it kinetic magnetoelectric effect:}
$M_a
\propto\tau K_{ba}E_b$,

\end{itemize}
where ${\bf j}$ is the induced current density, ${\bf M}$ is the
induced magnetisation, ${\bf E}$ or ${\bf E}(\omega)$ is the amplitude
of the static or optical electric field, $\tau$ is the relaxation
time of the conduction electrons, and $\varepsilon_{abc}$ is the alternating tensor.
The reader is referred to Ref.~\onlinecite{tsirkin-prb18} for more
details such as the prefactors in the expressions above, as well as
the formulas for the current-induced Faraday effect and natural
optical activity, both of which are also implemented in the \module{gyrotropic} module.

\subsection{postw90.x: Spin Hall conductivity}
The spin Hall effect (SHE) is a phenomenon in which a  
spin current is generated by applying an electric field. The current is often
transverse to the field (Hall-like), but this is not always the case.\cite{wimmer-prb15} 
The SHE is characterised by the spin Hall conductivity (SHC) tensor $\sigma_{ab}^{\text{spin},c}$ as follows:
\begin{equation}
J_a^{\text{spin},c}(\omega)= \sigma_{ab}^{\text{spin},c}(\omega)
E_b(\omega),
\end{equation}    
where 
$J_a^{\text{spin},c}$ is the spin-current density along direction $a$ with its spin pointing along $c$, and $E_b$ is the external electric field of frequency $\omega$ applied along $b$. 
In non-magnetic materials
the equal number of up- and down-spin electrons
forces the AHE to vanish,
resulting in a pure spin current.

Like the AHC, the SHC contains
both
intrinsic and extrinsic contributions.\cite{RevModPhys.87.1213} 
The intrinsic contribution to the SHC can be calculated from the following Kubo formula,\cite{PhysRevB.98.214402} 
\begin{subequations}\label{eq:shc_kubo}
\begin{align}
&\sigma_{ab}^{\text{spin},c}(\omega) = 
-\frac{e^2}{\hbar}\frac{1}{V N}\sum_{\bfk}
\sum_{n}
f_{n\bfk} \Omega_{n\bfk,ab}^{\text{spin},c}(\omega),\label{eq:shc_kubo_1}\\
&\Omega_{n\bfk,ab}^{\text{spin},c}(\omega) = {\hbar}^2 \sum_{
	m\ne n}\frac{-2\operatorname{Im}[\langle \psi_{n\bfk}|
	\frac{2}{\hbar}{j}_a^{\text{spin},c}|\psi_{m\bfk}\rangle
	\langle\psi_{m\bfk}|{v}_b|\psi_{n\bfk}\rangle]}
{(\epsilon_{n\bfk}-\epsilon_{m\bfk})^2-(\hbar\omega+i\eta)^2},\label{eq:shc_kubo_2}
\end{align}
\end{subequations}
where ${s}_c$, ${v}_a$ and ${j}_a^{\text{spin},c}=\frac{1}{2}\{{s}_c,{v}_a\}$
 are the  spin, velocity and spin current 
operators, respectively; 
$V$ is the cell volume, and $N$ is the total number of $k$-points used to sample the BZ.
Equations~\eqref{eq:shc_kubo} are very similar to the Kubo formula for the AHC, except for the 
replacement of a velocity matrix element by 
a spin-current matrix element.
As mentioned in the previous two subsections,
Wannier-interpolation techniques are very efficient at calculating such quantities. 

A Wannier-interpolation method scheme for evaluating the intrinsic SHC was developed in
Ref.~\onlinecite{PhysRevB.98.214402} (see also Ref.~\onlinecite{ryoo-prb19} for a related but independent work). 
The required 
quantities from the underlying {\it ab initio}
calculation
are 
the spin matrix elements $S_{mn\bfk,a}^{(0)} = \langle \psi_{m\bfk}^{(0)} | {s}_{a} | 
\psi_{n\bfk}^{(0)} \rangle$, 
the Hamiltonian matrix elements 
$H_{mn\bfk}^{(0)} = \langle \psi_{m\bfk}^{(0)} | {H} | 
\psi_{n\bfk}^{(0)} \rangle=\epsilon^{(0)}_{m\bf k}\delta_{mn}$, 
and the overlap matrix elements of \Eqref{eq:def-Mmn}.
Since the calculation of all these quantities has been previously implemented in \pwtowan{} (the interface code between \pwscf{}
and \Wannier{}), 
this advantageous interpolation scheme can be readily used while
keeping to a minimum
the interaction between the \textit{ab initio} code and {\sc Wannier90}. 

\begin{figure}
\includegraphics[width=8cm]{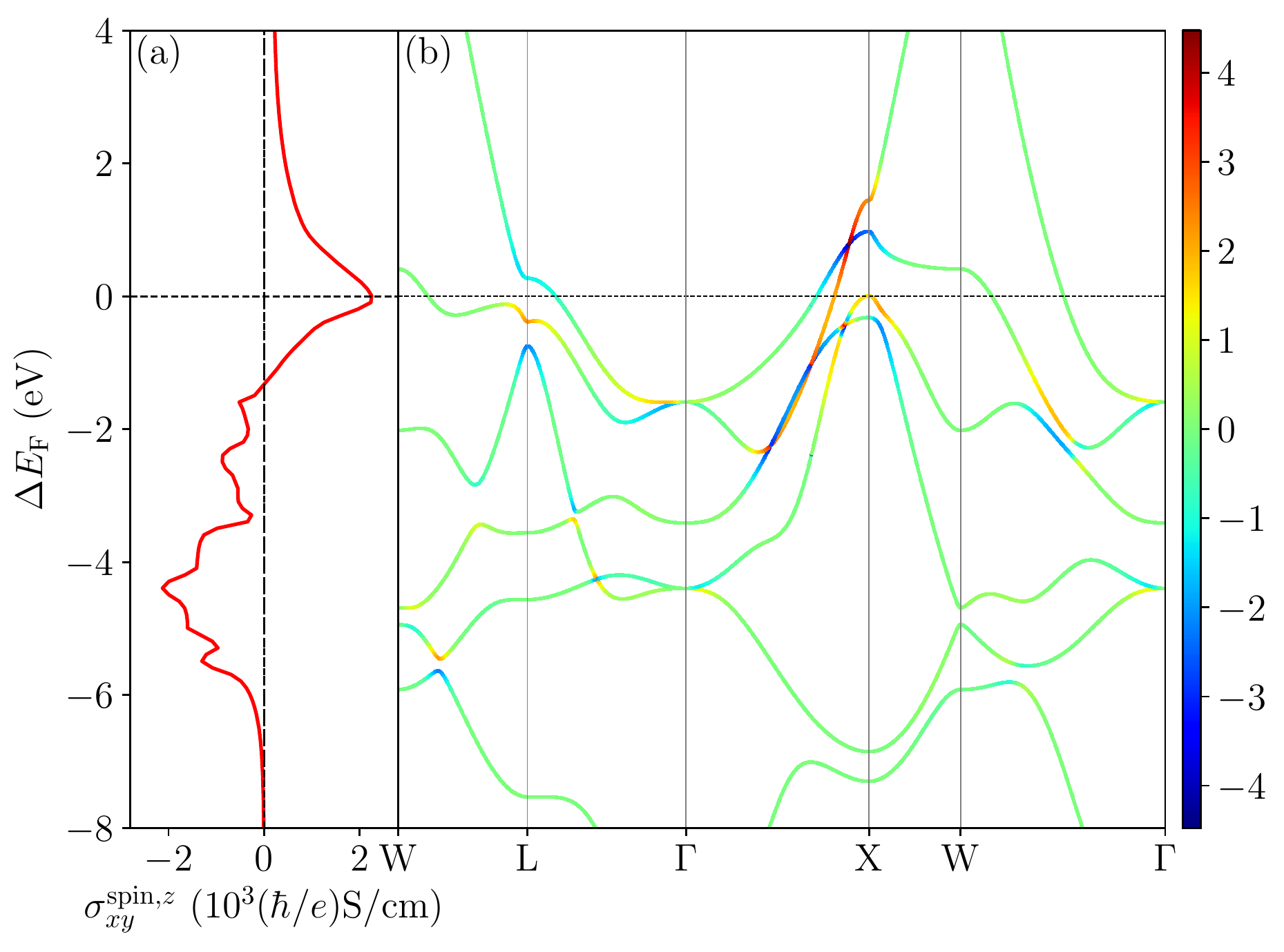}
\caption{\label{fig:pt_shc} 
(a) Intrinsic spin Hall conductivity $\sigma_{xy}^{\text{spin},z}$ of fcc Pt, plotted  
as a function of the shift in Fermi energy relative to its self-consistent value. 
(b) Band structure colour-coded as a function of the value of the spin Berry curvature $\Omega_{n\bfk,xy}^{\text{spin},z}$ of Eq.\eqref{eq:shc_kubo_2}. Note that the value plotted in the graph is $r(\Omega_{n\bfk,xy}^{\text{spin},z})$, with the function $r(x)$ defined by $r(x) = \left\{\begin{array}{ll}
x/10 & |x| < 10 \\
\log_{10}(|x|)\mathrm{sign}(x) & |x| \geq 10
\end{array}\right.$, with $\mathrm{sign}$ being the sign function and $\log_{10}$ the base-10 logarithm.}
\end{figure}

The application of the method to fcc Pt is illustrated in Fig.~\ref{fig:pt_shc}. Panel~(a) shows the calculated SHC as a function of the Fermi-level
position, and panel~(b) depicts the ``spin Berry curvature'' of Eq.~\eqref{eq:shc_kubo_2} that gives the contribution from each band state to the SHC.
The aforementioned functionalities have been incorporated in the \module{berry}, \module{kpath} and \module{kslice} modules of \exec{postw90}.

\subsection{postw90.x: Parallelisation improvements}

The original implementation of the \module{berry} module in \exec{postw90} (for computing Berry-phase properties such as orbital magnetisation and anomalous Hall conductivity\cite{lopez-prb12}), introduced in \Wannier{} v2.0, was written with code readability in mind and had not been optimised for computational speed. In \Wannier{} v3.0, all parts of the \module{berry} module have been parallelised while keeping the code readable; moreover, its scalability has been improved, accelerating its performance by several orders of magnitude.\cite{whitepaper}

To illustrate the improvements in performance we present calculations on a 128-atom supercell of GaAs interstitially doped with Mn. We use a lattice constant of the elementary cell of 5.65~\AA.
We use norm-conserving relativistic pseudopotentials with the PBE
exchange-correlation functional. The energy cut-off for the plane waves
is set to 40~Ry, and the Brillouin-zone sampling of the supercell is $3\times 3\times 3$.
We use a Gaussian metallic smearing with a broadening of 0.015 Ry. 
For the non-self-consistent step of the calculation, 600 bands are computed and used to construct 517 Wannier functions. The initial projections are chosen as a set of $sp^3$ orbitals centred on each Ga and As atom, and a set of $d$ orbitals on Mn. The calculations were performed on the Prometheus supercomputer of PL-GRID (in Poland). 

The Berry-phase calculations can be performed in three distinct ways: 
(i) 3D quantities in $k$-space 
(routine \inputvar{berry\_main}), (ii) the same quantities resolved on 2D planes (routine \module{kslice}), 
and (iii) 1D paths (routine \module{kpath}) in the Brillouin zone. 
In the benchmarks, we will refer to these three cases as ``Berry 3D'', ``Berry 2D'', and ``Berry 1D'', respectively.

The first optimisation target was the function {\tt utility\_rotate} in the module \module{utility},
which calculates a matrix product of the form $B = R^{\dagger}AR$ using Fortran's 
built-in {\tt matmul} function. The new routine {\tt utility$\_$rotate$\_$new}
uses instead BLAS and performs about 5.7 times better than the original one, giving
a total speedup for {\tt berry$\_$main} of about 55$\%$.

A second performance-critical section of code was identified in the routine {\tt get\_imfgh\_k\_list}, 
which took more than 50$\%$ of the total run-time of {\tt berry\_main}. 
This routine computes three quantities: $F_{\alpha\beta}$, $G_{\alpha\beta}$ and $H_{\alpha\beta}$,
which are defined in Eqs. (51), (66) and (56) of Ref.~\onlinecite{lopez-prb12}. By some algebraic 
transformations, it was possible to reduce 25 calls to {\tt matmul}, carried out in the innermost
runtime-critical loop, to only 5 calls. After replacement of {\tt matmul} with the Basic Linear Algebra Subprogram (BLAS), the speed up
of this routine exceeds a factor of 11, and the total time spent in {\tt berry\_main} is 2.5 
times shorter (including the speed-up from the first optimisation).

In the third step, a bottleneck was eliminated in the initialisation phase, where 
{\tt mpi\_bcast} was waiting more than two minutes for the master rank to broadcast 
the parameters. The majority of this time was spent in loops computing matrix products of the form 
$S = (V_1)^{\dagger}S_0V_2$. Again, we replaced this with two calls to the BLAS {\tt gemm} routine. This resulted in a speed-up of a factor of 610 for the calculation 
of this matrix product in our test case, and the total initialisation time dropped to less than 15 seconds. 
In total, the {\tt berry\_main} routine runs about 5 times faster than it did originally.

\begin{figure}
    \centering
    \includegraphics[width=6cm]{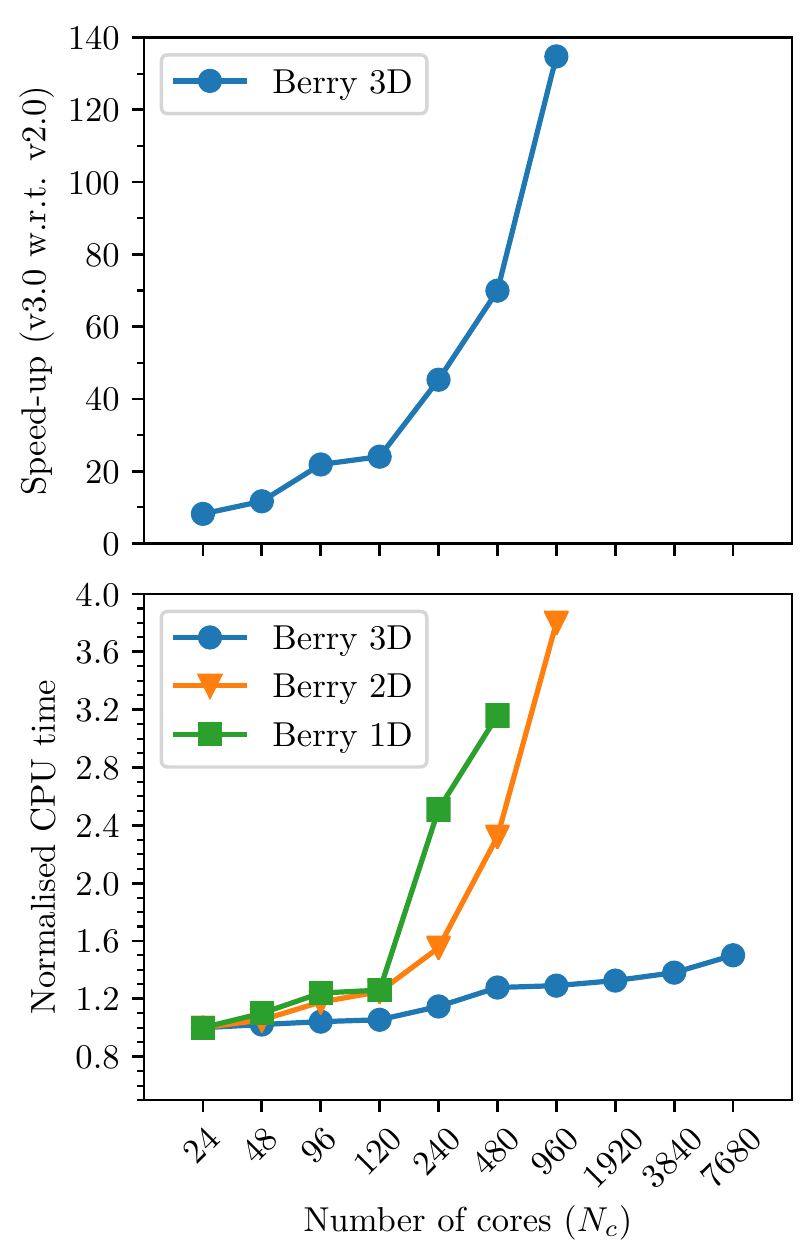}
    \caption{(Top) Speedup of the new \Wannier{} v3.0 with respect to v2.0, for a run of the berry module (mode ``Berry 3D'') on the test system described in the text, demonstrating the improvements implemented in the new version of the code. (Bottom) Total CPU time (defined as total walltime times number of CPUs) for the three cases ``Berry 3D'', ``Berry 2D'' and ``Berry 1D'', normalised with respect to the same case run with $N_{\text{cpu}}=24$, for the \Wannier{} v3.0 code. Note that calculations with $N_{\text{cpu}}\ge 480$ for ``Berry 3D'' were run on a denser grid ($100\times 100\times 100$ rather than $30\times 30\times 30$) and values have been rescaled using the time measured for both grids at $N_{\text{cpu}}= 480$.}
    \label{fig:berry-scalability}
\end{figure}

Finally, the routines \module{kslice} and \module{kpath} were parallelised. 
The scalability results of {\tt berry\_main}, \module{kslice} and \module{kpath} are presented in 
Fig.~\ref{fig:berry-scalability},
and a comparison with the scalability of the previous version of {\sc berry\_main} is also given. Absolute times for some of the calculations are reported in Table~\ref{tab:times-berry}.

\begin{table}[hb]
  \centering
  \begin{tabular}{ccccc}
  \hline
  Mode & $k$-grid & $N_{\text{c}}$ & Time (s) \\
  \hline
  \multicolumn{4}{c}{version 3.0} \\
  \hline
  \multirow{5}{*}{Berry 3D} & 30$\times$30$\times$30 & 24 &  6903 \\
        & 30$\times$30$\times$30 & 48 &  3527 \\
        & 30$\times$30$\times$30 & 480 &  441 \\
        & 100$\times$100$\times$100 & 480 &  13041 \\
        & 100$\times$100$\times$100 & 7680 &  957 \\
  Berry 2D & 100$\times$100 & 24 & 1389 \\
  Berry 1D & 10000 & 24 & 12639 \\
  \hline
  \multicolumn{4}{c}{version 2.0} \\
  \hline
  \multirow{2}{*}{Berry 3D} & 30$\times$30$\times$30 & 24 &  56497 \\
        & 30$\times$30$\times$30 & 48 &  40279 \\
  \hline
  \end{tabular}
  \caption{\label{tab:times-berry}Wall-time for some of the runs performed with the Berry module, before (\Wannier{} v2.0) and after (\Wannier{} v3.0) the optimisations, for the test system described in the main text. $N_{\text{c}}$ indicates the number of cores used in the calculation.}
  \end{table}

\subsection{GW bands interpolation}
While density-functional theory (DFT) is the method of choice for most applications in materials modelling, it is well known that DFT is not meant to provide spectral properties such as band structures, band gaps and optical spectra.
Green's function formulation of many-body perturbation theory (MBPT) \cite{fetter} overcomes this limitation, and allows the excitation spectrum to be obtained from the knowledge of the Green's function. Within MBPT the interacting electronic Green's function $G(\mathbf{r},\mathbf{r'},\omega)$ may be expressed in terms of the non-interacting Green's function $G^0(\mathbf{r},\mathbf{r'},\omega)$ and the so-called self-energy $\Sigma(\mathbf{r},\mathbf{r'},\omega)$, where several accurate approximations for $\Sigma$ have been developed and implemented into first-principles codes\cite{martin_reining_ceperley_2016}.  While maximally-localised Wannier functions for self-consistent GW quasiparticles have been discussed in Ref.~\onlinecite{hamann_gw_09}, here we focus on the protocol to perform bands interpolation at the one-shot G$_0$W$_0$ level. For solids, the G$_0$W$_0$ approximation has proven to be an excellent compromise between accuracy and computational cost and it has become the most popular MBPT technique in computational materials science \cite{lucia_review_18}. In the standard one-shot G$_0$W$_0$ approach, $\Sigma$ is written in terms of the Kohn--Sham (KS) Green's function and the RPA dielectric matrix, both obtained from the knowledge of DFT-KS orbitals and eigenenergies. Quasi-particle (QP) energies are obtained from:
\begin{equation}
\epsilon_{n\bfk}^{\rm QP} = \epsilon_{n\bfk} + Z_{n\bfk} \braket{\psi_{n\bfk}|\Sigma(\epsilon_{n\bfk})-V_{\rm xc}|\psi_{n\bfk}},
\end{equation}
where  $\psi_{n\bfk}$ and $\epsilon_{n\bfk}$ are the KS orbitals and eigenenergies, $Z_{n\bfk}$ is the so-called renormalisation factor and $V_{\rm xc}$ is the DFT exchange-correlation potential. In addition, in the standard G$_0$W$_0$ approximation the QP orbitals are approximated by the KS orbitals.
At variance with DFT, QP corrections for a given $k$-point require knowledge of the KS
orbitals and eigenenergies at all points $(\bfk + \mathbf{q})$ in reciprocal space. In practice, codes such as Yambo~\cite{yambo_paper_09} compute QP corrections on a regular grid and rely on interpolation schemes to obtain the full band structure along high-symmetry lines. 
\Wannier{} supports the use of G$_0$W$_0$ QP corrections through the general interface \texttt{gw2wannier90.py} distributed with \Wannier{}, while dedicated tools for \QE{} and Yambo allow an efficient use of symmetries. Here we briefly outline the procedure for performing Wannier interpolation at the G$_0$W$_0$ level using \QE{} and Yambo.
The procedure starts by obtaining MLWFs at the DFT level. While \Wannier{} works with uniform coarse meshes on the full BZ (FBZ), Yambo uses symmetries to compute quantities on the irreducible BZ (IBZ). In addition, the G$_0$W$_0$ self-energy may require finer $k$-point grids to achieve convergence compared to those required for the charge density in DFT or the Wannier interpolation itself.
The \texttt{k-mapper.py} utility allows the user to quickly select only the symmetry-inequivalent $k$-points in the IBZ that belong to the grid used by \Wannier{}. At this point, Yambo computes the QP corrections on these selected $k$-points only. After that, a post-processing code of Yambo (\texttt{ypp}) unfolds the QP corrections onto the full BZ as required by \Wannier{}. Using the unfolded QP corrections, the utility \texttt{gw2wannier90.py} corrects and reorders in energy both the KS eigenvalues and the input matrices. After reading these eigenvalues and matrices, \Wannier{} can proceed as usual to interpolate the desired quantities such as the band structure, but now at the G$_0$W$_0$ level.

\section{Automatic Wannier functions: the SCDM method}\label{sec:scdm}
An alternative method for generating localised Wannier functions, known as the selected columns of the density matrix (SCDM) algorithm, has been proposed by Damle, Lin and Ying \cite{DL_2015_SCDM,DL_2018_SIAM}. At its core the scheme exploits the information stored in the real-space representation of the single-particle density matrix, a gauge-invariant quantity. Localisation of the resulting functions is a direct consequence of the well-known nearsightedness principle\cite{Kohn_PRB7,Prodan_PNAS_2005} of electronic structure in extended systems with a gapped Hamiltonian, \ie{}, insulators and semiconductors. 
In these cases, the density matrix is exponentially localised along the off-diagonal direction in its real-space representation $\rho(\bfr,\bfrp)$ and it is generally accepted that Wannier functions with an exponential decay also exist; numerical studies have confirmed this claim for a number of  materials, and there exist formal proofs for multiband \mbox{time-reversal-invariant}  insulators\cite{Brouder_PRL_2007,He_PRL_86,Fiorenza2016}. Since the SCDM method does not minimise a given gauge-dependent localisation measure via a minimisation procedure, it is free from any issue regarding the dependence on initial conditions, \ie{}, it does not require a good initial guess of localised orbitals. It also
avoids other problems associated with a minimisation procedure, such as getting stuck in local minima. More generally, the localised Wannier functions provided by the SCDM method can be used as starting points for the MLWF minimisation procedure, by using them to generate the $\matAk$ projection matrices needed by \Wannier{}.

For extended insulating systems, the density matrix is given by 
\begin{equation}
	\rho = \sum_{\bfk} P_{\bfk} = \sum_{n=1}^{J}\sum_{\bfk} \kpsi{n}{\bfk}\bpsi{n}{\bfk}.
\end{equation}
As shown in Sec.~\ref{sec:background}, the $P_{\bfk}$ are the spectral projectors associated with the crystal Hamiltonian operator $H_\bfk$ onto the valence space $\mathcal{S}_\bfk$, hence their rank is $N_{\rm e}$ (number of valence  electrons). Moreover, they are analytic functions of $\bfk$  
and also manifestly gauge invariant\cite{Nenciu_RMP_63,Panati_CMP_2013}. As mentioned above, the nearsightedness principle\cite{Prodan_PNAS_2005} guarantees that the columns of the kernels \mbox{$P_{\bfk}(\bfr,\bfrp)=\elm{\bfr}{P_{\bfk}}{\bfrp}$}
are localised along the off-diagonal direction and therefore they may be used to construct a localised basis. 
If we consider a discretisation of the $J$ Bloch states at each $\bfk$ on a real-space grid of $N_{\rm g}$ points, we can arrange the wavefunctions into the columns of a unitary $N_{\rm g}\times J$ $k$-dependent matrix $\Psi_{\bfk}$
\begin{equation}
\Psi_{\bfk} = \begin{pmatrix} \psi_{1\bfk}(\bfr_1) & \dots & \psi_{J\bfk}(\bfr_1) \\ 
 \vdots & \ddots & \vdots \\
 \psi_{1\bfk}(\bfr_{N_{\rm g}}) & \dots & \psi_{J\bfk}(\bfr_{N_{\rm g}}) \end{pmatrix},
\end{equation}
such that $P_{\bfk,ij} = \left(\Psi_{\bfk}\Psi_{\bfk}^\dag\right)\phantom{}_{ij}$ is a $N_{\rm g}{\times}N_{\rm g}$ matrix. In this representation, it is  straightforward to see that the columns of $P_{\bfk}(\bfr_i,\bfr_j)$ are projections of extremely localised functions (\ie{}, Dirac-delta functions localised on the grid points) onto the valence eigenspace. 
As a result, selecting any linearly-independent
subset of $J$ of them will yield a localised basis for the span of $P(\bfr,\bfrp)$. However, randomly selecting $J$ columns does not guarantee that a well-conditioned basis will be obtained. For instance, there could be too much
overlap between the selected columns.
Conceptually, the most well conditioned columns may be found via a QR factorisation with column pivoting (QRCP) applied to $P(\bfr,\bfrp)$, in the form $P\Pi = QR$, with $\Pi$ being a matrix permuting the columns of $P$, $Q$ a unitary matrix and $R$
an upper-triangular matrix (not to be confused with the lattice vector ${\bf R}$, or with the matrix $R^{({\bf k},{\bf b})}$ defined in Eq.~\eqref{eq:Rmn}, or with the shift vector of Eq.~\eqref{eq:shift-vector}), and where $\Pi$ is chosen so that $|R_{11}| \ge |R_{22}| \ge \cdots \ge |R_{nn}|$.
Then the $J$ columns forming a localised basis set are chosen to be the first $J$ of the matrix with permuted columns $P\Pi$.

The SCDM-$k$\cite{DL_2018_SIAM} method suggests that it is sufficient to apply the QRCP factorisation at $\bfk=\boldsymbol{0}$ ($\Gamma$ point) only, and use the same selection of columns at all $k$-points. However, this is still often impractical since $P_{\bGamma}$ is prohibitively expensive to construct and store in memory. Therefore an alternative procedure is proposed, for which the columns can be computed via the QRCP of the (smaller) matrix $\Psi_{\bGamma}^\dagger$ instead:
\begin{equation}
\Psi_{\bGamma}^\dag \Pi = Q'R',
\end{equation}
i.e., the same $\Pi$ matrix is obtained by computing a QRCP on $\Psi^\dagger$ only. 
Once the set of columns has been obtained, we need to impose the orthonormality constraint on the chosen columns without destroying their locality in real space. This can be achieved by a L\"owdin orthogonalisation, similarly to Eq.~\eqref{eq:initial-U}.
In particular, the selection of columns of $\Psi_{\bGamma}$ can be used to select the columns of all $\Psi_{\bfk}$, which in turn define the $\amn$ matrices needed as input by \Wannier{} to start the MLWF minimisation procedure, by defining \mbox{$\amn = \psi^\ast_{m\bfk} (\bfr_{\Pi(n)})$}, where the $\Pi(n)$ is the index of the $n^{\rm th}$ column of $P$ after permutation with $\Pi$. In fact, we can write the $n^{\rm th}$ column of $P$ after permutation, $P_{\bfk}(\bfr,\bfr_{\Pi(n)})$, as
\begin{eqnarray}
P_{\bfk}(\bfr,\bfr_{\Pi(n)}) & = &\sum_{m=1}^{J} \psi_{m\bfk}(\bfr)\psi^\ast_{m\bfk}(\bfr_{\Pi(n)}) \\
&= &\phi_{\Pi(n),\bfk} \equiv \sum_{m=1}^{J} \psi_{m\bfk}(\bfr)\amn.\label{eq:SCDM-Vmn}
\end{eqnarray}
The unitary matrix $\matUk$ sought for is then constructed via L\"owdin orthogonalisation
\begin{equation}
\matUk = \matAk(\matAk^{\dag}\matAk)^{-\nicefrac{1}{2}} = \matAk \matSk^{-\nicefrac{1}{2}}.
\end{equation}
We can also extend the SCDM-$k$ method to the case where the Bloch states are represented as two-component spinor wavefunctions $\psi_{n\mathbf{k}}(\mathbf{r},\alpha)$, e.g., when including spin-orbit interaction in the Hamiltonian. Here, $\alpha = \uparrow, \downarrow$ is the spinor index.
In this case, we include the spin index as well as the position index to perform QRCP.
First, we define the $2N_g \times J$ matrix $\Psi_{\mathbf{k}}$
\begin{equation} \label{eq:scdm_spinor} 
    \Psi_{\mathbf{k}} = 
  \begin{pmatrix}
    \psi_{1\mathbf{k}}(\mathbf{r}_1,\uparrow) & \dots &\psi_{J\mathbf{k}}(\mathbf{r}_1,\uparrow) \\
    \psi_{1\mathbf{k}}(\mathbf{r}_1,\downarrow) & \dots &\psi_{J\mathbf{k}}(\mathbf{r}_1,\downarrow) \\
    \vdots & \ddots & \vdots\\
    \psi_{1\mathbf{k}}(\mathbf{r}_{N_g},\uparrow) & \dots & \psi_{J\mathbf{k}}(\mathbf{r}_{N_g},\uparrow) \\
    \psi_{1\mathbf{k}}(\mathbf{r}_{N_g},\downarrow) & \dots & \psi_{J\mathbf{k}}(\mathbf{r}_{N_g},\downarrow)
  \end{pmatrix}.
\end{equation}
Next, as in the spinless case, the QRCP of $\Psi_{\mathbf{\Gamma}}^\dagger$ is computed, and the first $J$ columns of the $\Pi$ matrix are selected.
Now, $\Pi(n)$, the index of the $n^{\rm th}$ column of $P$ after permutation with $\Pi$, determines both the position index $\mathbf{r}_{\Pi(n)}$ and the spin index $\alpha_{\Pi(n)}$.
We define $A_{mn\mathbf{k}} = \psi^*_{m\mathbf{k}}(\mathbf{r}_{\Pi(n)}, \alpha_{\Pi(n)})$ and perform L\"owdin orthogonalisation to obtain the unitary matrix $U_{\mathbf{k}}$.

In the case of entangled bands, we need to introduce a so-called quasi-density matrix defined as
\begin{equation}
P_{\bfk} = \sum_n \ket{\psi_{n\bfk}}f(\epsilon_{n\bfk})\bra{\psi_{n\bfk}},\label{eq:gen_Pk}
\end{equation}
where $f(\epsilon_{n\bfk}) \in [0,1]$ is a generalisation of the Fermi-Dirac probability for the occupied states. Also in this case we only use the information at $\Gamma$ to generate the permutation matrix.
Depending on what kind of entangled manifold one is interested in, $f(\epsilon)$ can be modelled with various functional forms. In particular, the authors of Ref.~\onlinecite{DL_2018_SIAM} suggest the following three forms: 
\begin{enumerate}
	\item Isolated manifold, \eg{}, the valence bands of an insulator or a semiconductor: $f(\epsilon)$ is a step function, with the step inside the energy gap $\Delta\epsilon_{\rm g} = \epsilon_{\rm c} - \epsilon_{\rm v}$, where $\epsilon_{\rm c(v)}$ represents the minimum (maximum) of the conduction (valence) band: \begin{equation}f(\epsilon)=\theta(\epsilon_{\rm v}+\Delta\epsilon_{\rm g}/2 - \epsilon).\end{equation}
	Both $\Delta\epsilon_{\rm g}$ and $\epsilon_{\rm v}$ are not free parameters, as they may be obtained directly from the {\it ab initio} calculation.
	\item Entangled manifold (case I), \eg{}, the valence bands and low-lying conduction bands in a semiconductor: $f(\epsilon)$ is a complementary error function: \begin{equation}f(\epsilon)=\frac{1}{2}\mathrm{erfc}\left(\frac{\epsilon - \mu}{\sigma}\right),\end{equation}
	where $\mu$ is used to shift the mid-value of the complementary error function, so that states with energy equal to $\mu$ have a weight of $f(\mu)=1/2$. The parameter $\sigma$ is used to gauge the ``broadness'' of the distribution function.
	\item Entangled manifold (case II), \eg{}, the $d$ bands in a transition metal: $f(\epsilon)$ is a Gaussian function \begin{equation}f(\epsilon) = \exp\left(-\frac{(\epsilon - \mu)^2}{\sigma^2}\right).\end{equation}
\end{enumerate}
The procedure then follows as in the previous case, by computing a QRCP factorisation on the quasi-density matrix. It is worth to note that in the case of an entangled manifold, the SCDM method requires the selection of two real numbers: $\mu$ and $\sigma$, as well as the number of Wannier functions to disentangle $J$. These parameters play a crucial role in the selection of the columns of the density matrix. While the selection of these parameters requires some care, as a rule of thumb (e.g., in entangled case I) $\sigma$ is of the order $2-5$~eV (which is the energy range of a typical bandwidth), while $\mu$ can often be set around the Fermi energy (but the exact value depends on various factors, including the number $J$ of bands chosen and the specific properties of the bands of interest).
It is worth to mention that since the SCDM-$k$ method is employed as an alternative way of specifying a set of initial projections and hence to compute the $\matAk$ matrices in Eq.~\eqref{eq:initial-U}, the disentanglement procedure can be used in exactly the same way as described in Sec.~\ref{sec:entangled}. However, in the case of entangled bands the column selection is done on a quasi-density matrix, which implicitly defines a working subspace larger than the target subspace of dimension $J$. We find that for well-known systems SCDM-$k$ is typically already capable of selecting a smooth manifold and no further subspace selection is needed.

This method is now implemented as part of the \pwtowan{} interface code to \textsc{Quantum ESPRESSO}. We have decided to implement the algorithm in the interface code(s) rather than in \Wannier{} itself, because the SCDM method requires knowledge of the wavefunctions $\psi_{n\bfk}$, which are only available in the {\it ab initio} code.

In \Wannier{} only a single new input parameter \inputvar{auto\_projections} is required. This disables the check on the number of projections specified in the input file (as we rely on SCDM to provide us with the initial guesses) and adds a new entry to the \verb|<seedname>.nnkp| file (which is read by \pwtowan{} in order to compute the quantities required by \Wannier{}) that specifies the number of Wannier functions required. The remaining control parameters for the SCDM method are specified in the input file for the \pwtowan{} code, including whether to use the SCDM method, the functional form of the $f(\epsilon)$ function in Eq.~\eqref{eq:gen_Pk} and, optionally, the values of $\mu$ and $\sigma$ in the definition of $f(\epsilon)$.

\section{\label{sec:aiida}Automation and workflows: \aiidawan{} plugin}

AiiDA~\cite{pizzi-AiiDA} (Automated Interactive Infrastructure and Database for Computational Science) is an informatics infrastructure that helps researchers in managing, automating, storing and sharing their computations and results. AiiDA automatically tracks the entire provenance of every calculation to ensure full reproducibility, which is also stored in a tailored database for efficient querying of previous results. Moreover, it provides a workflow engine, allowing researchers to implement high-level workflows to automate sequences of tedious or complex calculation steps. AiiDA supports simulation codes via a plugin interface, and over 30 different plugins are available to date~\cite{aiida-registry}. 

Among these, the \aiidawan{} plugin provides support for the \Wannier{} code. Users interact with the code (to submit calculations and retrieve the results) via the high-level python interface provided by AiiDA rather than directly creating the \Wannier{} input files. AiiDA will then handle automatically the various steps involved in submitting calculations to a cluster computer, retrieving and storing the results, and parsing them into a database. Furthermore, using the AiiDA workflow system users can chain pre-processing and post-processing calculations automatically (e.g., the preliminary electronic structure calculation with an \emph{ab initio} code). These scientific workflows, moreover, can encode in a reproducible form the scientific knowledge of expert computational researchers in the field on how to run the simulations, choose the numerical parameters and recover from potential errors. In turn, their availability reduces the training time of new researchers, eliminates sources of error and enables large-scale high-throughput simulations.

The \aiidawan{} plugin expects that each calculation takes a few well-defined input parameters.
Among the most important ones, a \Wannier{} calculation run via AiiDA requires that the following input nodes are provided:
an input crystal \texttt{structure}, a node of
\texttt{parameters} with a dictionary of input flags for \Wannier{}, a node with the list of \texttt{kpoints}, a node representing the atomic \texttt{projections}, and a \texttt{local\_input\_folder} or \texttt{remote\_input\_folder} node containing the necessary input files (\texttt{.amn}, \texttt{.mmn}, \texttt{.nnkp}, \texttt{.eig}, \texttt{.dmn}) for the \Wannier{} calculation as generated by an \emph{ab initio} code.

All of these parameters, with the exception of projections, are generic to AiiDA to facilitate their reuse with different simulation codes. 
 More detailed information on all inputs can be found in the \aiidawan{} package documentation~\cite{aiidawan-docs}. 

After the \Wannier{} execution is completed, the \aiidawan{} plugin provides parsers that are able to detect whether the convergence was successful and retrieve key parameters including the centres of the Wannier functions and their spread, as well as the different components of the spread ($\omi$, $\omd$, $\omod$ and $\Omega$), and (if computed) the maximum imaginary/real ratio of the Wannier functions and the interpolated band structure. 

\begin{figure*}[tb]
  \includegraphics[width=16cm]{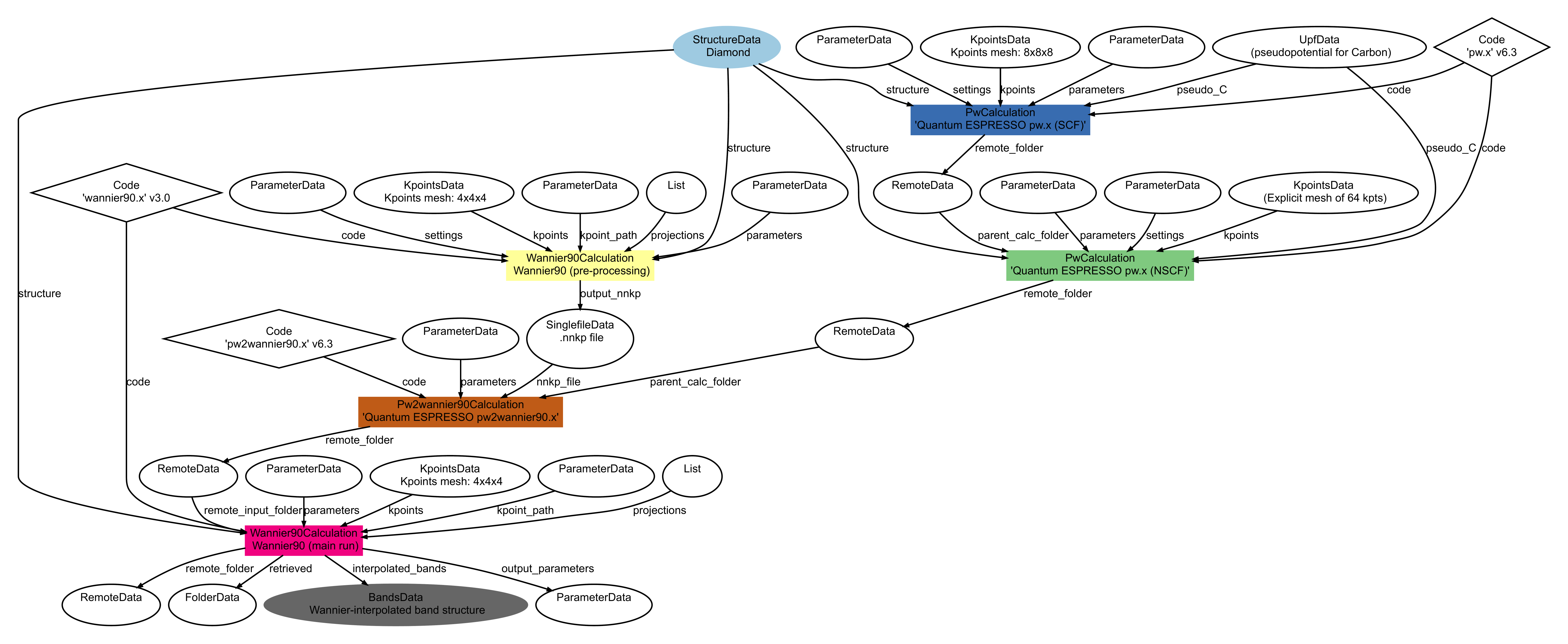}

\caption{\label{fig:aiida-graph}The provenance graph automatically generated by AiiDA when running a \Wannier{} calculation for a diamond crystal using \QE{} as the DFT code. Rectangles represent executions of calculations, ellipses represent data nodes, and diamonds are code executables. Graph edges connect calculations to their inputs and outputs. In particular, the following calculations are visible: Quantum ESPRESSO pw.x SCF (dark blue) and NSCF (green), \QE{} \pwtowan{} (brown), and \Wannier{} pre-processing (yellow) and minimisation run (purple). The initial diamond structure (light blue) and the final interpolated band structure (dark grey) are also highlighted.}
\end{figure*}

The whole simulation is stored in the form of a graph, representing explicitly the provenance of the data generated including
all inputs and outputs of the codes used in the workflow.
An example of a provenance graph, automatically generated by AiiDA when running a \QE{} calculation followed by a \Wannier{} calculation, is shown in Fig.~\ref{fig:aiida-graph}.

We emphasise that the availability of a platform to run \Wannier{} in a fully-automated high-throughput way via the \aiidawan{} plugin has already proved to be beneficial for the \Wannier{} code itself. Indeed, it has pushed the development of additional features or improvements now part of \Wannier{} v3.0, including additional output files to facilitate output parsing and improvements in some of the algorithms and their default parameters to increase robustness.

\section{\label{sec:goodcodepractices}Modern software engineering practices}
In this section, we describe a number of modern software engineering practices that
are now part of the development cycle of the \Wannier{} code.
In particular, \Wannier{} includes a number of tests that are run at every commit via a continuous integration approach, as well as nightly in a dedicated test farm. 
Version control is handled using git and the code is hosted on the GitHub platform~\cite{GitHub}. We follow the fork and pull-request model, in which users can duplicate (fork) the project into their own private repository, make their own changes, and make a pull request (i.e., request that their changes be incorporated back into the main repository). When a pull request is made, a series of tests are automatically performed: the test suite is run both in serial and parallel using the Travis continuous integration platform~\cite{travis}, and code coverage is checked using \package{codecov}~\cite{codecov}. If these tests are successful then the changes are reviewed by members of the \Wannier{} developers group and, if the code meets the published coding guidelines, it can be merged into the development branch.

In addition, while interaction with end users happens via a mailing-list forum, discussion among developers is now tracked using GitHub issues. This facilitates the maintenance of independent conversation threads for each different code issue, new feature proposal or bug. These can easily reference code lines as well as be referenced in code commit messages. Moreover, for every new bug report a new issue is opened, and pull requests that close the issue clearly refer to it. This approach facilitates tracking back the reasoning behind the changes in case a similar problem resurfaces.

In the remainder of this section we describe more in detail some of these modern software engineering practices.

\subsection{\label{sec:ford}Code documentation (FORD)}
The initial release of \Wannier{} came with extensive documentation in the form of a \emph{User Guide} describing the methodology, input flags to the program and format of the input and output files. This document was aimed at the end users running the software. Documentation of the code itself was done via standard code comments. In order to foster not only a community of users but also of code \emph{contributors} to \Wannier{}, we have now created an additional documentation of the internal structure of the code. This makes the code more approachable, particularly for new contributors. 
To create this code documentation in a fully automated fashion, we use the FORD (FORtran Documenter)\cite{ford-docs} documentation generator. We have chosen this over other existing documentation solutions because of FORD's specific support for Fortran. 
This tool parses the Fortran source, and generates a hyperlinked (HTML) index of source files, modules, procedures, types and programs defined in the code.  Furthermore, it constructs graphs showing the dependencies between different modules and subroutines. Additional information can be provided in the form of special in-code comments (marked with double exclamation marks) describing in more detail variables, modules or subroutines. By tightly coupling the code to its documentation using in-code comments, the documentation maintenance efforts are greatly reduced, decreasing the risk of having outdated documentation. 
The compiled version of the documentation for the most recent code version is made available on the Wannier90 website~\cite{aiidawan-ford-docs}.

\subsection{\label{sec:testing}Testing infrastructure and continuous integration}

With the recent opening to the community of the \Wannier{} development, it has become crucial to create a non-regression test suite to ensure that new developments do not break existing functionalities of the code. 
Its availability facilitates the maintenance of the code and ensures its long-term stability.

The \Wannier{} test suite relies on a modified version of James Spencer's python \texttt{testcode.py}~\cite{Spencer-testcode}. This provides the functionality to run tests and compare selected quantities parsed from the output files against benchmarked values. 

At present, the \Wannier{} test suite includes over 50 tests which are run both in serial and parallel and cover over 60\% of the source code (with many modules exceeding 80\% coverage). The code coverage is measured with the \package{codecov} software~\cite{codecov}. 
Developers are now required to add tests when adding new features to the code to ensure that their additions work as expected. This also ensures that future changes to the code will never break that functionality.
Two different test approaches are implemented, serving different purposes. 

First, the \Wannier{} repository is now linked with the Travis continuous integration platform~\cite{travis} to prevent introducing errors and bugs into the main code branch.
Upon any commit to the GitHub repository, 
the test suite is run both in serial and in parallel. Any test failure is reported back to the GitHub webpage. Additionally, for tests run against pull requests, any failed test results in the pull request being blocked and not permitted to merge. Contributors will first need to change their code to fix the problems highlighted in the tests; pull requests are able to be merged only after all tests pass successfully.

Second, nightly automatic tests are run on a \package{Buildbot} test-farm.
The test-farm compiles and runs the code with a combination of compilers and libraries (current compilers include GFortran v6.4.0 and v7.3.0, Intel Fortran Compiler v17 and v18, and PGI compiler v18.05; current MPI libraries include Open MPI v1.10.7 and v3.1.3, Intel MPI v17 and MVAPICH v2.3b).
This ensures that the code runs correctly on various high-performance computer (HPC) architectures. More information on the test-farm can be found on the \Wannier{} GitHub wiki website\cite{github-wiki}.

In addition to these tests, we have implemented git pre-commit hooks to help keep the same code style in all source files. The current pre-commit hooks run Patrick Seewald's Fortran source code formatter \package{fprettify}~\cite{fprettify} to remove trailing whitespaces at the end of a line and to enforce a consistent indentation style. These precommit hooks, besides validating the code, can reformat it automatically. Developers may simply run the formatter code to convert the source to a valid format.
If a developer installs the pre-commit hooks, these will be run automatically before every commit. Even if this is not the case, these tests are also run on Travis; therefore, a pull request that does not conform to the standard code style cannot be merged before the style is fixed.

\subsection{Command-line interface and dry-run}
The command-line interface of the code has been improved.
Just running \exec{wannier90} without parameters shows a short explanation of the available command line options. In addition, a \texttt{-v} flag has been added to print the version of the code, as well as a new \texttt{-d} dry-run mode, that just parses the input file to perform all needed checks of the inputs without running the actual calculation. The latter functionality is particularly useful to be used in input validators for \Wannier{} or to precalculate quantities computed by the code at the beginning of the simulation (such as nearest-neighbour shells, $b$-vectors or expected memory usage) and use this information to validate the run or optimise it (e.g., to decide the parallelisation strategy within automated AiiDA workflows).

\subsection{Library mode}
\Wannier{} also comes with a library mode, where the core code functionality can be compiled into a library that can then be linked by external programs. This library mode is used as the default interaction protocol by some interface codes. The library mode provides only support for a subset of the full functionality, in particular at the moment it only supports serial execution.
We have now added and improved support for the use of excluded bands also within the library mode. Moreover, beside supporting the generation of a statically-linked library, we now also support the generation of dynamically-linked versions. Finally, we have added a minimal test code, run together with all other tests in the test suite, that serves both to verify that the library functionality works as expected, and as an example of the interface of the library mode.

\section{\label{sec:conclusions}Conclusions and outlook}
\Wannier{} v2.0 was released in October 2013 with a small update for v2.1 in January 2017. The results and developments of the past years, presented in this work, were released in \Wannier{} v3.0 in February 2019. Thanks to the transition of \Wannier{} to a community code, 
\Wannier{} includes now a large number of new functionalities and improvements that make it very robust, efficient and rich with features. These include the implementation of new methods for the calculation of WFs and for the generation of the initial projections; parallelisation and optimisations; interfaces with new codes, methods and infrastructures; new user functionality; improved documentation; and various bug fixes. The effect of enlarging the community of developers is not only visible in the large number of contributions to the code, 
but also in the modern software engineering practices that we have put in place, that help improve the robustness and reliability of the code and facilitate its maintenance by the core \Wannier{} developers group and its long-term sustainability.

The next major improvement that we are planning is the implementation of a more robust and general library mode. The features that we envision are: (1) the possibility to call the code from C or Fortran codes without the need to store files but by passing all variables from memory; (2) a more general library interface that is easily extensible in the future when new functionality is added; and (3) the possibility to run \Wannier{} from a parallel MPI code, both by running each instance in parallel and by allowing massively-parallel codes to call, in parallel, various instances of \Wannier{} on various structures or with different parameters. This improvement will demand a significant restructuring of most of the codebase and requires a good design of the new interface. Currently we are drafting the new library interface, by collecting feedback and use cases from the various contributors and users of the code, to ensure that the new library mode can be beneficial to all different possible use cases.

\acknowledgements

We acknowledge code contributions by
Daniel Aberg (\myfont{w90pov} code), Lampros Andrinopoulos (\myfont{w90vdw} code), Pablo Aguado Puente (\module{gyrotropic} module), Raffaello Bianco ($k$-slice plotting), 
Marco Buongiorno Nardelli ({\tt dosqc} v1.0 subroutines upon which some of \module{transport} is based), Stefano de Gironcoli (\pwtowan{} interface to \QE{}), Pablo Garcia Fernandez (matrix elements of the position operator), Nicholas Hine (\myfont{w90vdw} code), Young-Su Lee (specialised $\Gamma$-point routines and transport), Antoine Levitt (preconditioning), Graham Lopez (extension of \pwtowan{} to add terms needed for orbital magnetisation), Radu Miron (constrained centres), Nicolas Poilvert (transport routines), Michel Posternak (original plotting routines), Rei Sakuma (symmetry-adapted Wannier functions), Gabriele Sclauzero (disentanglement in spheres in $k$-space), Matthew Shelley (transport routines), Christian Stieger (routine to print the $U$ matrices), David Strubbe (various bug fixes and improvements), 
Timo Thonhauser (extension of \pwtowan{} to add terms needed for orbital magnetisation), 
as well as the participants of the first \Wannier{} developers meeting in San Sebasti\'an (Spain) in 2016 for useful discussions (Daniel Fritsch, Victor Garcia Suarez, Pablo Garcia Fernandez, Jan-Philipp Hanke, Ji Hoon Ryoo, J\"urg Hutter, Javier Junquera, Liang Liang, Michael Obermeyer, Gianluca Prandini, Christian Stieger, Paolo Umari).
The WDG acknowledges financial support from the NCCR MARVEL of the Swiss National Science Foundation, the European Union's Centre of Excellence E-CAM (grant no.~676531), and the Thomas Young Centre for Theory and Simulation of Materials for support (grant no.~TYC-101).

\end{document}